\documentclass[aps,prb,twocolumn,epsfig,pdflatex]{revtex4-1}
\UseRawInputEncoding
\usepackage{amsmath,amssymb}
\usepackage{graphicx}
\usepackage{epsfig}
\usepackage{epstopdf}

\usepackage{float}
\usepackage[usenames,dvipsnames]{xcolor}
\usepackage{amsthm,comment}
\usepackage{booktabs}
\usepackage[export]{adjustbox}
\usepackage[section]{placeins}
\usepackage[caption=false]{subfig}
\usepackage[percent]{overpic}
\bibpunct{[}{]}{,}{n}{}{}
\definecolor{red1}{HTML}{FF4136}
\definecolor{green1}{HTML}{00802b}

\usepackage[hidelinks]{hyperref} 

\hypersetup{colorlinks=true,citecolor=Red,linkcolor=Blue,urlcolor=Blue}
\usepackage[export]{adjustbox}

\graphicspath{{Figures/}}

\begin{document}

\title {Quench dynamics of two component dipolar fermions \\
subject to  a quasiperiodic potential }

\author{Bradraj Pandey$^{1,2}$,   Elbio Dagotto$^{1,2}$ and Swapan K. Pati$^{3}$}

\affiliation{$^1$Department of Physics and Astronomy, University of Tennessee, Knoxville, Tennessee 37996, USA \\
$^2$Materials Science and Technology Division, Oak Ridge National Laboratory, Oak Ridge, Tennessee 37831, USA \\
$^{3}$
Theoretical Sciences Unit, Jawaharlal Nehru Centre for Advanced Scientific Research, 
Jakkur Campus, Bangalore 560064, India
 }

\date{\today}

\begin{abstract}
Motivated by recent experiments in fermionic polar gases, we study the non-equilibrium dynamics of 
two-component dipolar fermions subject to a quasiperiodic potential.  
We investigate the localization of charge and spin degrees of freedom time evolving with a long-range spin-SU(2)
symmetric fermionic Hamiltonian, by calculating experimentally accessible dynamical observables.  
To study the non-equilibrium dynamics, we start the time evolution with two initial states at half-filling:
(i) product state with doublons $|\uparrow \downarrow 0 \uparrow \downarrow 0 \uparrow \downarrow 0 \uparrow \downarrow 0 \uparrow \downarrow \rangle$
and (ii) product state with singlons $|\uparrow \ \downarrow  
\ \uparrow \  \downarrow \  \uparrow \  \downarrow \  \uparrow \  \downarrow \ 
\uparrow \  \downarrow \ \rangle$. We carried out the real-time evolution via
the fermionic Hamiltonian using exact diagonalization(ED) and the time-dependent variational principle (TDVP) for finite
Matrix product states(MPSs), within experimentally relevant time scales. For the product state with doublons, we observe a 
delocalized to localized phase transition varying disorder strengths, by monitoring  the decay of charge imbalance with time. 
For the long-range interacting Hamiltonian of our focus, and in the presence of strong enough disorder, starting the time evolution with singlons 
we find  a strong reduction in the spin delocalization, contrary to results of previous studies using the disordered 
short-range (on-site) Hubbard model with SU(2) symmetry. 
In addition to the quench dynamics, we also demonstrate the localization of charge and spin in the full energy spectrum of the long-range spin-SU(2) 
symmetric Hamiltonian, by monitoring spin and charge auto-correlation functions. Our predictions for localization of both charge and spin should be observable in ultra-cold experiments with fermionic dipolar atoms subject to a quasiperiodic potential.

\end{abstract}


\maketitle

\section{Introduction}
Recent progress in ultra-cold atomic gases experiments provides a promising platform for
studying the non-equilibrium dynamics of interacting systems~\cite{Jeisert,altman}. 
More specifically, the nearly perfect isolation from the environment, and tunability of ultra-cold atomic gases
properties, allow experimentalists to probe the 
unitary time evolution of isolated quantum systems~\cite{Langen}. 
These isolated systems can evolve in time through a thermalization-like process, 
where the memory of the initial state is lost because of its
own dynamics~\cite{deut,sred,Greiner,rigol}. 
Remarkably, the system itself acts effectively as a heat bath, via the exchange of energy and particles 
among portions of the sample~\cite{huse}. We refer to these systems as ergodic. However, there are exceptions. 
Both integrable and localized systems fail to thermalize, preserving the 
memory of the initial states~\cite{kon,ander}. Recently, localized many-body systems 
with disorder potential have attracted considerable attention due to its possible
 applications in quantum information theory and for exploring new paradigms of 
statistical mechanics~\cite{sapien}. 

The localization of non-interacting particles  
in the presence of quenched disorder was first described by Anderson~\cite{ander} [Anderson localization (AL)]. In this case the system shows both an absence of transport and  absence of thermalization. In low-dimensional 
disordered systems, the presence of interactions gives rise to a 
many-body localization (MBL) phase~\cite{dmbasko,ivgor,vogan,apal}. Interestingly, this 
MBL phase  shows very different characteristics in terms of spectral and 
dynamical properties in comparison to the ergodic systems~\cite{rnand}. 
In the presence of sufficiently strong disorder, many-body localization occurs in the full energy-spectrum,
leading to the area-law scaling of excited eigenstates~\cite{bauer} (whereas ergodic systems exhibit a volume law scaling~\cite{vidmar}),
and can be described by a complete set of quasi-local integrals of motion, similarly as integrable systems~\cite{papic,ogan}.
In the case of many-body localization, the energy-level spacing statistics follows a Poisson distribution, 
whereas ergodic systems follow a Wigner-Dyson distribution~\cite{vogan}.

 The most popular way to characterize  the MBL phase
is to follow the real-time dynamics of some initial product state in the
presence of both disorder and interactions. In a MBL phase, the system carries 
the memory of the initial state even at infinite time~\cite{mapapic,muller}.
 Under the unitary-time evolution from an initial state in presence of strong enough disorder, 
the local observables evolve following a power-law decay,
but they converge to nonvanishing stationary values~\cite{Dabanin,fiemini}, while in an ergodic system they would
converge to zero with increasing time.
 The transport of energy, charge, and spins are absent in both the AL and MBL phases~\cite{Falet}, 
while in the ergodic phase a fast transfer of energy and particles after a sudden quench is observed~\cite{hkim}. 

The dynamics of entanglement entropy provides an important tool to differentiate between AL, MBL, and ergodic 
phases~\cite{muller,Falet,Dabanin,fiemini,alex}.
For short-range interacting disordered systems, in the MBL phase the entanglement entropy 
increases logarithmically with time~\cite{prosen,jbard,mserb}, in the AL phase this entanglement 
entropy saturates to a small but finite value after a short time, 
and finally in the ergodic phase the entanglement grows linearly with time (ballistically)~\cite{hkim}. 
Interestingly, the MBL phase also has been found in long-range interacting spin systems~\cite{jsmith,laumann} 
(or in spinless fermions~\cite{nag}), where entanglement entropy was found to grow algebraically with time, namely $t^{\eta}$~\cite{mpino,masella}.
The power law exponent $\eta$ takes a universal value $\eta_c=0.33$ at the delocalized to localized MBL transition for 
one component long-range interacting systems ~\cite{masella}. 

\subsection{Previous results using the short range (on-site) disordered Hubbard model}

Experiments in cold atomic systems in the presence of a quasiperiodic potential, for on-site (i.e. short-range) fermionic systems 
support quite well the theoretical and numerical predictions for the MBL phase with regards to the localization of charge~\cite{mschre,pbor,lusch}. 
Specifically, in Ref.~\cite{mschre} the many-body localization of interacting fermionic systems 
in a quasiperiodic potential has been shown by observing the decay of the density imbalance with time. 
The system was modeled in terms of the on-site interacting fermionic Hubbard model and the time evolution was carried out starting 
with charge density wave states. But the localized vs delocalized character of the spin sector was not addressed.
Thus, the evidence provided is only of partial MBL.

Interestingly, most of the experiments described above were performed with spinfull fermionic Hubbard models,
where the system has a continuous non-Abelian SU(2) symmetry. However, theoretically it has been suggested 
 that in the presence of such a continuous non-Abelian symmetry the system can not 
have a full MBL phase even at large disorder~\cite{protop,potter,prakash}.  
In fact, numerical studies of the one-dimensional disordered Hubbard model with on-site interactions 
(and with disorder in the charge sector)
showed that the charge degree of freedom is localized but the spin degree of freedom is delocalized, contrary
to the properties of the MBL where both charge and spin should be localized~\cite{prelov,zakr,koza}.
 In Ref.~\cite{koza},  by mapping the Hubbard model to an effective spin Heisenberg model in the presence of large disorder,
sub-diffusive transport in the spin channel was observed. 
In a recent study~\cite{ivan}, it was argued that the delocalization of spin can also lead to
delocalization of charge after a very long time, depending upon the fermionic occupancy, rendering the system
even more different than the MBL state. 
The restoration of the full MBL phase was demonstrated in Ref.~\cite{prelov,sroda,rubem}, 
by {\it breaking} the spin SU(2) symmetry which now leads to the localization of the spin, together with the charge.
Thus, the presence or absence of many-body localization in SU(2) symmetric systems is
 an ongoing topic of research~\cite{Iadeco}, but the theoretical consensus is that on-site (short-range) Hubbard models with SU(2) symmetry do not produce a MBL phase and we need to break SU(2) to stabilize such a state.

\subsection{New results: SU(2) symmetric fermionic model with long-range interactions and disorder}

The above described studies of the MBL phase were based on the disordered Hubbard model using {\it local on-site} interactions. In this publication, for the first time we explore the localization of spin and charge, using a continuous spin-SU(2) symmetric interacting fermionic Hamiltonian including {\it long-range} interactions.
This long-range interacting Hamiltonian is quite relevant for experimental studies of
 non-equilibrium dynamics in polar Fermi gases~\cite{kkni,cheng,barun}, trapped ions~\cite{Jurcevic}, and  Rydberg gases~\cite{zeiher}, because they naturally have long-range interactions.
However, as  argued in Refs.~\cite{protop,panda}, the continuous non-Abelian symmetry of the disordered Hamiltonian creates degeneracy in the spectrum. 
For these reasons, a number of questions arise in the context of dynamics of charge and spin degree of freedom with spin-SU(2) 
symmetric long-range interacting disordered Hamiltonian. How does the long-range interaction influence the dynamics of 
charge and spin degrees of freedom? What will be the nature of entanglement entropy 
with time in the presence of strong disorder? Along with answering these questions, we also discuss the role of the initial state to 
observe localization of charge and spin in quench experiments with spin-SU(2) symmetric system.

As explained above, our work is motivated by recent experimental progress in the study
of polar Fermi gases~\cite{kkni,cheng,barun} and by experiments addressing the MBL with fermionic systems in quasiperiodic potentials~\cite{mschre,lusch}.
To explore the non-equilibrium dynamics, we follow the quench protocol, 
namely we start the time evolution with two simple and experimentally feasible product states, specifically the
{\it doublons} state $|\uparrow \downarrow 0 
\uparrow \downarrow 0 \uparrow \downarrow 0 \uparrow \downarrow 0 
\uparrow \downarrow ..\rangle$ and the {\it singlons} state $|\uparrow \ \downarrow \ 
\uparrow \  \downarrow \  \uparrow \  \downarrow \  \uparrow \  \downarrow \ 
\uparrow \  \downarrow \  ..\rangle$ at half filling\cite{bauer2}: 
\noindent
(a) Starting the time evolution 
from the product state with doublons, we demonstrate the transition from a delocalized phase to a 
localized phase by varying the disorder strength.
Analyzing the dynamics of local observables, such as the charge imbalance, and also the global observable entanglement, 
as well as charge and spin fluctuations, 
we are able to show the localization of both the spin and charge separately.
\noindent
(b) Starting the time evolution from the product state with singlons, in the presence of strong disorder
we discuss the effect of the long-range interaction on the dynamics corresponding to the spin degree of freedom, 
both in the large and small interaction limits. Assuming the charge is localized, we present a comparative study of the spin dynamics corresponding 
to the disordered on-site Hubbard model and our disordered long-range Hubbard model with SU(2) symmetry.
In contrast to the previous studies using the disordered local on-site interacting Hubbard model, here we find that the delocalization of 
the spin degree of freedom is {\it strongly reduced} for the long-range interacting model, even without breaking the spin-SU(2) symmetry.
More specifically, when the long-range and on-site couplings of our model are of comparable strength, we show that the
the spin imbalance decays very slowly and the entanglement entropy grows linearly with time.
Furthermore, we demonstrate the effect of long-range interactions on the localization of spin and charge in the full spectrum
by calculating the charge and spin auto correlation.

 The manuscript is organized as follows. In Sec.~II, we
discuss the long-range model Hamiltonian and method used. Section~III contains the results, and associated discussion, 
and it is divided into three subsections A, B and C. In subsection A, for  the product states with doublons we show
the presence of  a delocalized to localized transition
in the presence of a quasiperiodic potential.
In subsection B, we show the localization of the spin starting the time evolution from the product state with singlons,
 again in the presence of a quasiperiodic potential. In subsection C, we discuss the localization of the spin and charge in the full energy spectrum . Section~IV contains our conclusions.
In the Appendix, we discuss the stability of doublons for the on-site and long-range Hubbard models without disorder.
 Then, we provide an intuitive description for the
localization of the spin degree of freedom in terms of exchange interactions with a large disorder potential.
We have presented the long-time dynamics of charge and spin degrees of freedom for product states with doublons and singlons.

\section{Model and Methods}

In this study we consider two-component (pseudo spin-1/2) dipolar fermions in a one-dimensional 
lattice at half-filling.
The effective Hamiltonian is,
\begin{eqnarray} 
H= t_{hop} \sum_{i, \sigma}  \left (c^{\dagger}_{i,\sigma} c_{i+1,\sigma} +H.c. \right) + 
U\sum_i \hat{n}_{i,\uparrow} \hat{n}_{i,\downarrow} \nonumber \\ 
+ \sum_{\langle i \ne j \rangle} V(i,j)\tilde{n}_{i} \tilde{n}_{j}+ \Delta \sum_{i,\sigma} 
\hat{n}_i cos(2\pi \beta i+ \phi), 
\end{eqnarray}
\noindent where  $c_{i, \sigma}$ is the annihilation operator with spin $\sigma= \uparrow,\downarrow $
at site $i$. Here the $\uparrow$ and $\downarrow$ states refer to two hyperfine states 
of dipolar atoms or molecules. $\tilde{n}=(\hat{n}-\langle n \rangle)$ 
where $\hat{n}$ is the number operator and $\langle n \rangle$ is the average density, 
which is taken as $1$ because it is a half-filled system.
The hopping amplitude is denoted as $t_{hop}$ 
(we set $t_{hop}=1$ for our numerical calculation) and $U$ is the on-site interaction term. $V(i,j)$
is the long-range interaction potential, which depends on the relative orientation and distance
between the polarized dipoles as~\cite{barun,laumann,pandey} $V_{dd}\propto (1-3 \cos^2(\theta))/r^3$ in a standard
sphericall coordinates notation.
In an optical lattice by using the Feshbach resonance or by changing the lattice depth,
the on-site interaction strength $U$ can be modulated.
The model in Eq.(1) preserves both $U(1)$ and spin $SU(2)$  symmetries, 
related to the conservation of the total particle number $N$  
 and to the total spin of the system, respectively~\cite{jagar,japar}.

To study the effect of disorder on the localization of the spin and charge
degrees of freedom when using our long-range Hamiltonian, we added  a quasi-periodic 
on-site potential~\cite{mtez,shi}. The strength of disorder is 
measured by the coefficient $\Delta$ of this quasiperiodic potential and we choose 
$\beta =(\sqrt{5}-1)/2$, a commonly used value as employed in Refs.~\cite{mtez,shi}. The non-interacting model 
($U=V=0$ and $t_{hop}=1$) with quasiperiodic potential
shows a metal to insulator transition at $\Delta=2.0$~\cite{kohmoto}. Interestingly, for the 
critical disorder at $\Delta_C=2.0$ 
all the single particles state become localized for irrational values of $\beta$~\cite{mschre,kohmoto,aubre}.

The equilibrium phase diagram of this model (with $\Delta=0$) is well known~\cite{Mosade}.
For positive $U$ and $V$ (where $V$ is the prefactor of the dipolar interaction $1/{r^3}$), and in the range $U>2V$, the large on-site repulsive interaction leads the fermions 
to form a spin-density wave phase.
On the other hand, for large repulsive interactions $V$ and in the range $U<2V$,
two fermions tend to reside at the same site and form a charge density wave~\cite{6Zhang}.
Here, in order to study the non-equilibrium dynamics, we initialize the 
system in these two extreme kinds of product states: ({\it i}) a product state of doublons and empty sites, and ({\it ii}) a product state of singlons, both at overall half filling.
 
We have investigated numerically the time evolution of these initial product states $|\Psi(0)\rangle$,
 under the influence of the system  Hamiltonian, using $|\Psi(t)\rangle=\exp{\left(-iHt\right)}|\psi(0)\rangle$.
In particular, we used the time dependent 
exact-diagonalization (ED) method for system sizes $L=8$, $10$, and $12$. For larger
systems (up to  $L=64$ sites), using the ITensor library~\cite{itensor} we have implemented 
the time-dependent variational principle (TDVP)  
for the finite Matrix Product States (MPS) time-evolution method~\cite{Cirac,lubich}. 
To enlarge the bond dimension of the MPS for the TDVP time evolution, we have used the recently developed global subspace expansion algorithm~\cite{white,itensor}
for a few initial steps of the one-site TDVP1 time evolution (see Appendix~C for more details). 

\section{Results and discussion}
\subsection{Quench from the product state with doublons and empty sites at U=2V}
\begin{figure}[h]
\centering
\rotatebox{0}{\includegraphics*[width=\linewidth]{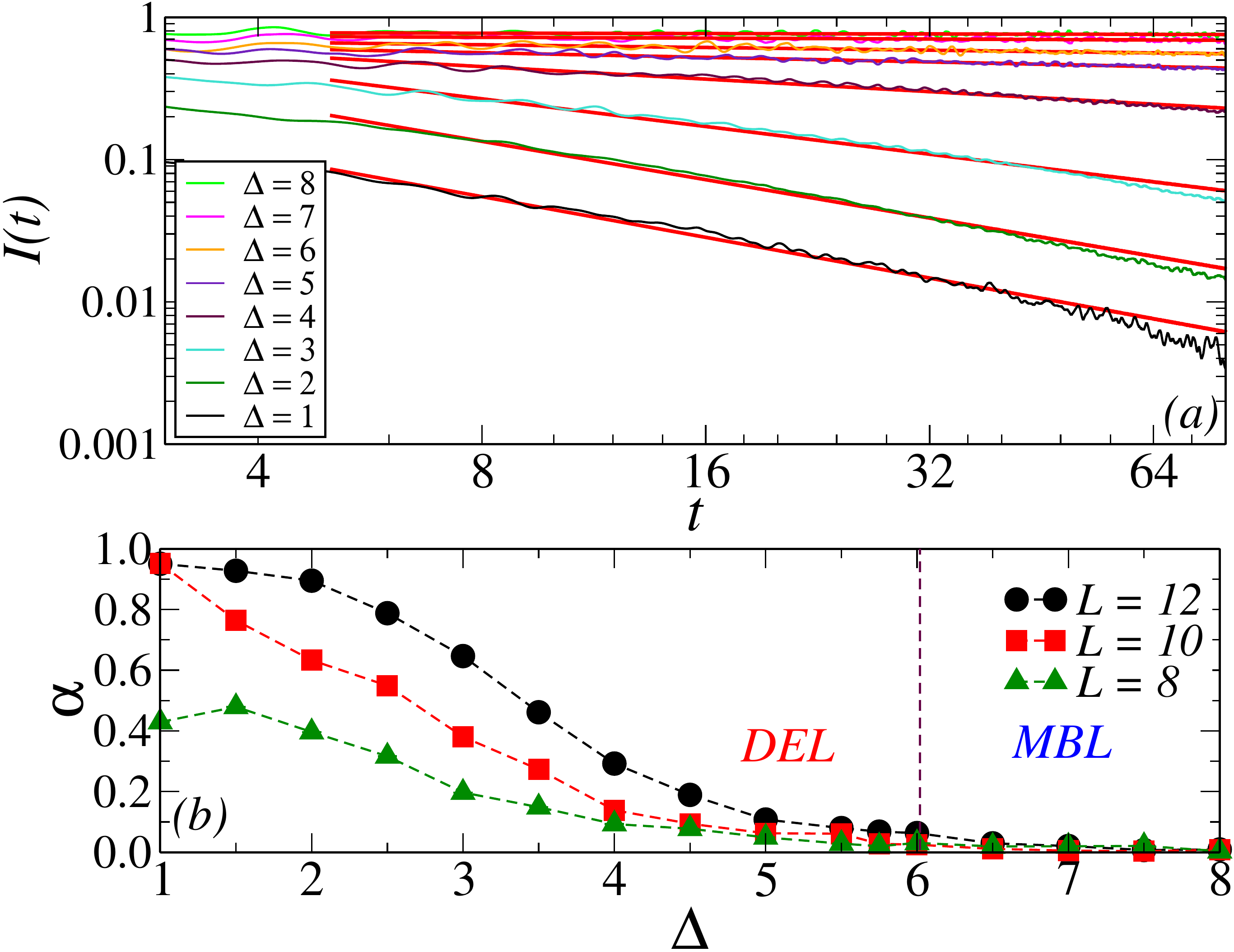}}
\caption{(a) Time evolution of the average charge imbalance $I(t)$ vs time $t$ (in log-log scale and time in units of $\hbar/t_{hop}$) 
at fixed interaction couplings $U=2V=8$ and for the various disorder strengths $\Delta$ shown in the inset. The red curves between $5t$ to $80t$ 
represent power-law fits to the $I(t)$ decay. 
 (b) Relaxation exponent $\alpha$ of the charge imbalance decay  vs $\Delta$, for $L=8$,  $L=10$ and $L=12$, showing 
the delocalized to localized transition increasing $\Delta$. The size of data point symbols (circle, square, triangle) represent the
fitting error bars. The delocalized region is denoted as DEL.} 
\label{fig1}
\end{figure}

In this subsection we explore the localization of charge and spin for our long-range model 
subject to a quasi-periodic disordering potential, starting the time evolution from a product state
with doublons and empty sites, namely $|\uparrow \downarrow 0 \uparrow \downarrow 0 
\uparrow \downarrow 0 \uparrow \downarrow 0 \uparrow \downarrow \rangle$. 
For the Hamiltonian Eq.(1), without disorder i.e. $\Delta=0$, the doublons become 
unstable near the transition point $U\sim2V$, 
where they decay to single fermions (as explained in Appendix A).
The boundary $U=2V$ is chosen because: (i) for $U>2V$, doublons are very stable due to energy conservation i.e. having so many doubly-occupied sites considerably increases the energy with respect to the ground state but there are no energy-conserving channels to reduce the number of these doublons, while (ii) for $U<2V$ the ground state already has doublons thus there is no reason for the product state to be unstable. {\it Only at, or very close to 
the boundary between phases, the product state of doublons can be unstable [see Fig.~\ref{fig8}(a)]}.

\subsubsection{Delocalized to Localized Transition at U=2V=8}

We have carried out the time evolution using the exact-diagonalization method ($L=12$),
at fixed values of the interaction parameters $U=2V=8.0$, 
and for various values of the disorder strengths $\Delta$. 
The transition from the delocalized (DEL) to the localized phase, 
is a dynamical quantum phase transition~\cite{apal} and can be 
observed in experiments by monitoring the time evolution of local observables~\cite{mschre} and  
entanglements~\cite{alex}.
\begin{figure}[h]
\centering
\rotatebox{0}{\includegraphics*[width=\linewidth]{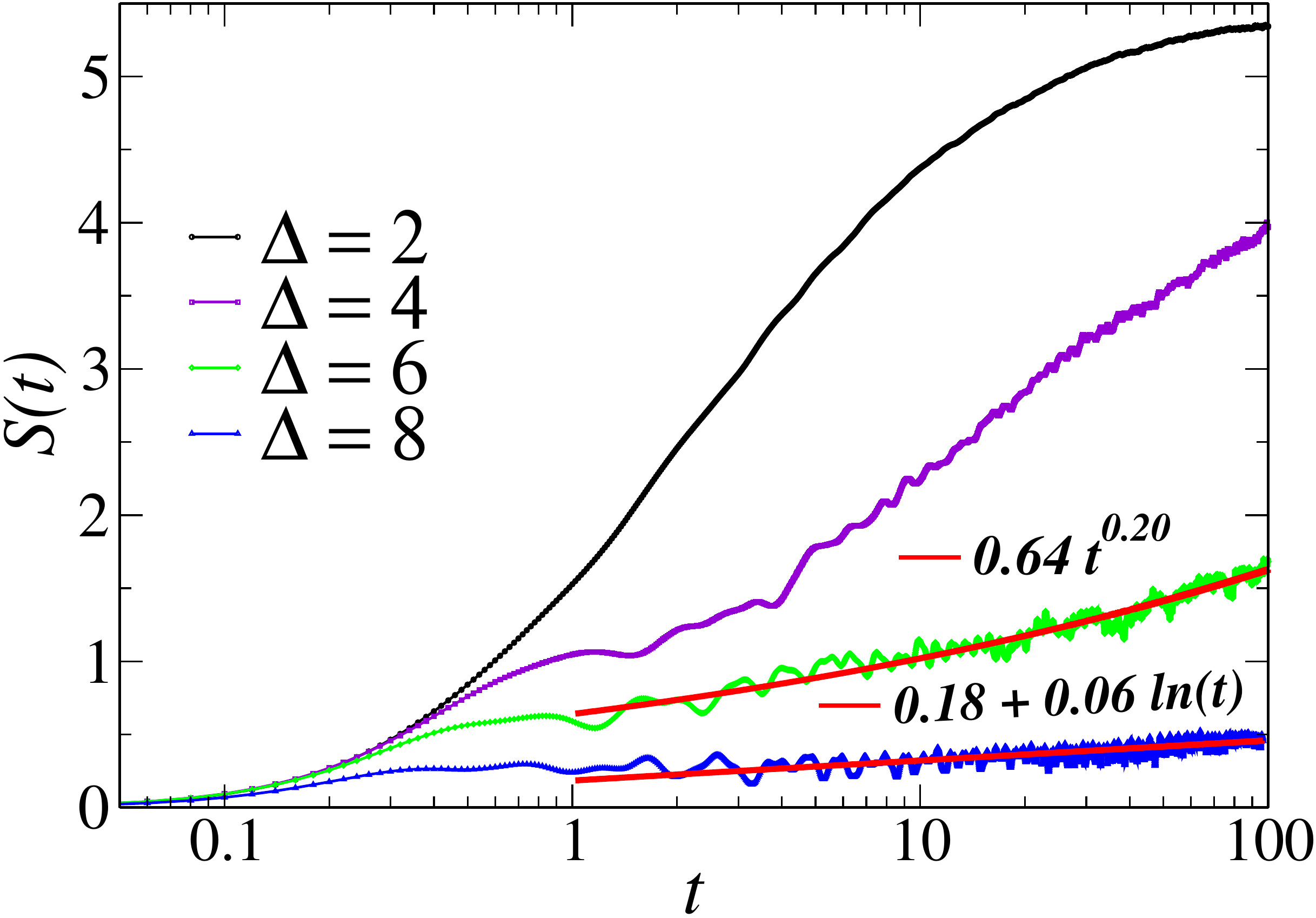}}
\caption{Time evolution of the entanglement entropy $S(t)$ with increasing time $t$ (in log scale and in units of $\hbar/t_{hop}$). We work at $U=2V=8$ and using several values of $\Delta$. The red curves show a power-law fit of
 $S(t)$ at $\Delta=6$ and a logarithmic fit at  $\Delta=8$.}
\label{fig2}
\end{figure}

In the MBL phase, the system carries a memory of the initial state and particles retain their initial
position even after long times. This has been observed in experiments 
by probing the time evolution of the charge imbalance $I(t) = \frac{N_e-N_o}{N_e+N_o}$,
where $N_e$ and $N_o$ is the number of fermions on even and odd sites, respectively~\cite{mschre}. 
For the charge-density wave state with doublons  $|\uparrow \downarrow 0 \uparrow \downarrow 0 
\uparrow \downarrow 0 \uparrow \downarrow 0 \uparrow \downarrow ..\rangle$,  at $t=0$ the charge 
imbalance $I(t)$ is one. 
Figure~\ref{fig1}(a) displays the time evolution of the average charge imbalance for various values of 
disorder strength $\Delta$.
Here, the charge imbalance $I(t)$ was averaged over  sites followed by averaging over 
ten different values of the phase factor $\phi$,
for each disorder strength $\Delta$. As shown in Fig.~\ref{fig1}(a), at low values of disorder, 
the imbalance decays  to zero (implying delocalization), while for large 
disorder strength, $I(t)$ saturates to a value close to the initial 1, suggesting localization of particles. 

To find more accurately the transition point from the delocalized regime to the localized phase, 
with increasing disorder strength, we have extracted the relaxation exponent $\alpha$ 
by power-law fitting the decay of $I(t)$ $\sim $ $t^{\alpha}$~\cite{lusch}.
We find that for low values of $\Delta$, the charge imbalance $I(t)$ decays sharply and with strong oscillation, 
giving comparatively a reduced time window to fit using a power law for $I(t)$. 
To achieve a better comparison of power-law fits across the delocalized to localized transition, 
we fit the data in the time window from $5t$ to $80t$ for all the values of $\Delta$ (Fig.~\ref{fig1}(a))~\cite{lusch}. 
Figure~\ref{fig1}(b) displays the relaxation exponent $\alpha$ vs disorder strength $\Delta$, for 
three different system sizes $L=8$, $L=10$, and $L=12$. 
The relaxation exponent $\alpha$ decreases monotonically in the delocalized phase with increasing  disorder strength $\Delta$.
Near the transition region ($5.5 \lesssim \Delta \lesssim 6.5$) this exponent $\alpha$ decreases slowly 
with increasing $\Delta$,
due to the slow dynamics near the delocalized to localized transition~\cite{lusch}. 
As shown in Fig.~\ref{fig1}(a), for $\Delta > 6.0$, $I(t)$ saturates to the initial value and 
 in Fig.~\ref{fig1}(b) the relaxation exponent $\alpha$ of different system sizes approaches 
very small values for $\Delta > 6.0$. 
We estimate the transition from delocalized to localized phase to be $\Delta_C \gtrsim 6.0 \pm 0.5$.
Interestingly, due to the strong on-site $U$ and long-range $V$ interactions,
 the transition from DEL to MBL occurred at values of $\Delta$ larger than those reported in recent experiments on fermionic on-site Hubbard model
 in a quasiperiodic potential~\cite{lusch}. 
\begin{figure}[h]
\centering
\rotatebox{0}{\includegraphics*[width=\linewidth]{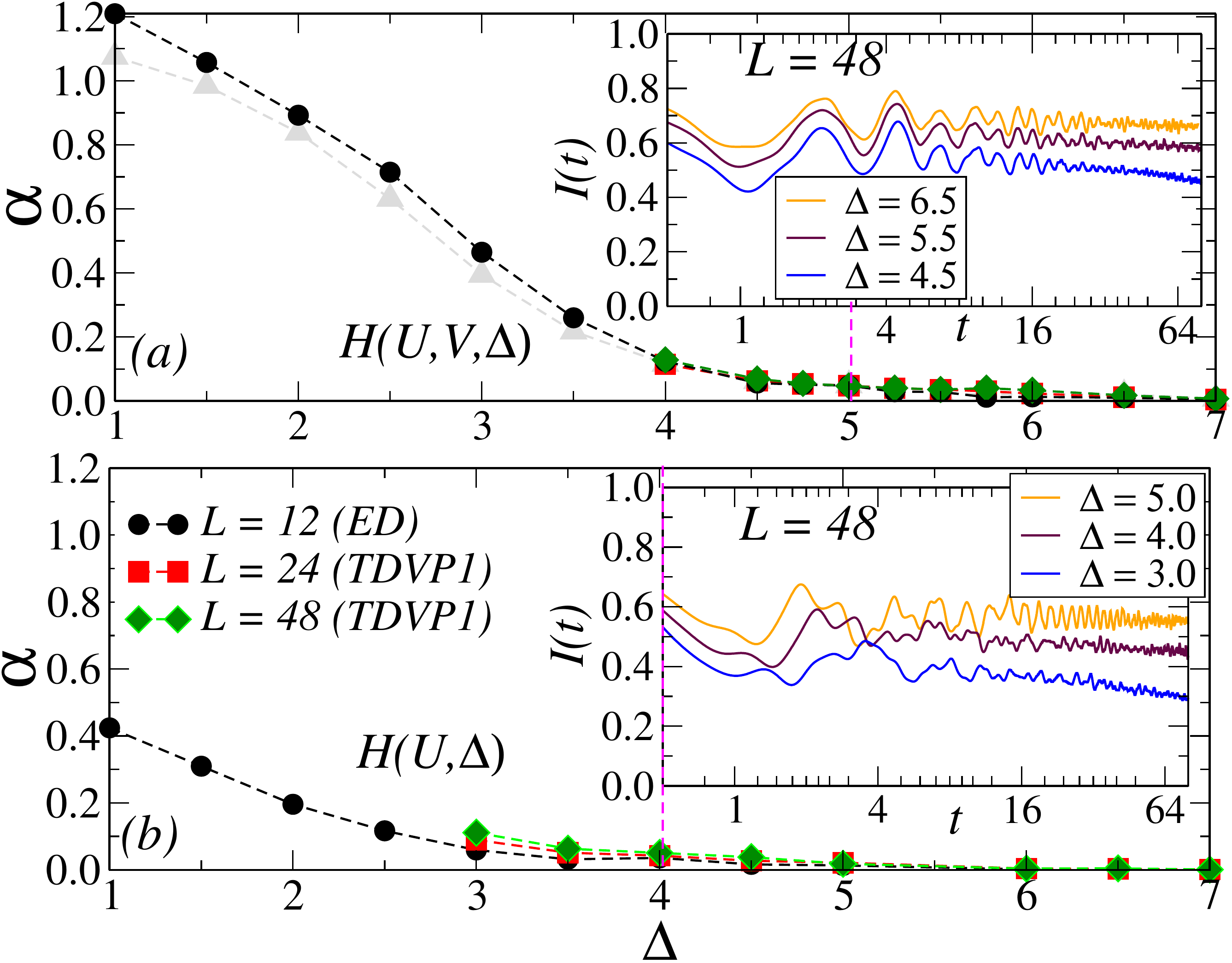}}
\caption{Comparison of decay exponent $\alpha$ obtained by power-law fit to the average charge imbalance $I(t)$ for 
(a)disordered extended-Hubbard [$H(U,V,\Delta)$] at $U=2V=4$ and (b) disordered Hubbard model [$H(U,\Delta)$]
at $U=4$. Grey color (triangle) data point symbol for long-range interacting Hamiltonian with same interaction parameter. 
Insets shows decay of charge imbalance vs time $t$(in log scale and in units of $\hbar/t_{hop}$) 
near the transition region of delocalize to localize transition.
These results were obtained by using TDVP1 for system sizes $L=24$ and $48$ and ED for $L=12$.} 
\label{fig3}
\end{figure}
In these experiments, the DEL to MBL transition was studied starting with a charge-density wave, 
where up and down spin atoms are randomly distributed on even lattice sites, and keeping all of the odd lattice sites empty~\cite{lusch}.

Figure~\ref{fig2} contains the time evolution of the entanglement entropy $S(t)=-Tr\left[ \rho_A(t) ln\rho_A(t)\right]$
for $L=10$, $U=2V=8$ and several values of the disorder strength $\Delta$, where $\rho_A(t)$ is the
reduced density matrix of block A of size $L/2$. 
At small $\Delta$, the entanglement entropy $S(t)$  grows rapidly with time
and saturates to large values, which is expected due to the breakdown of doublons
into single particles at $U=2V$. Near the phase transition, the entanglement entropy growth is 
power law with time ($S(t)\sim t^{\eta}$) with a small exponent $\eta=0.2$ (see Fig.~\ref{fig3}). 
The growth rate in the entanglement entropy further slows down with increasing $\Delta$. 
In this regime, the growth of the entanglement entropy $S(t)$ 
can be described by a logarithmic function of time $S(t)=0.06 ln(t)+0.18$, compatible with MBL.
Recently, evidence for the logarithmic growth of entanglement was reported experimentally in ultra-cold atomic systems
through the study of particle fluctuations and their correlations~\cite{alex}.    

\subsubsection{Delocalize to Localize Transition at U=2V=4}

Here we analyze the effect of $V$ on the localization of charge degrees of freedom, 
starting the time evolution with the product state with doublons. 
Using the recently developed time evolution method TDVP1 (see Appendix C for detail) near the  transition region,
we were able to calculate the charge imbalance $I(t)$ for significantly larger system sizes up to $L=48$ 
(typical system size used in cold atom experiments). For the TDVP time evolution,
we considered only nearest-neighbor $V$ terms along with the on-site interaction $U$.
This is a good approximation for larger values of disorder strength [see Fig.\ref{fig3}(a): the exponent $\alpha$ 
 obtained from long-range interaction $V$ coincides with the nearest-neighbor interaction $V$ for $\Delta \gtrsim 4.0$]. 
 The charge imbalance $I(t)$ shown in Fig.~\ref{fig3} is averaged over sites, 
followed by averaging over ten different values of phase factor $\phi$.

Figure~\ref{fig3}(a) displays the decay exponent $\alpha$ for the 
disordered extended-Hubbard model vs disorder strength $\Delta$.
The  exponent $\alpha$ obtained by power-law fit to the decay of charge imbalance $I(t)$ (as discussed in previous subsection)
for three different system sizes $L=12$ (ED), $24$, and $48$ (TDVP1). In the delocalized regime,
the exponent $\alpha$ monotonically decreases sharply with increase in $\Delta$. However,
near the delocalized to localized transition $\alpha$ decreases very slowly [$\Delta \gtrsim 4.5$].
In the localized region the decay exponents $\alpha$ shows a weak dependence on system size 
and $\alpha $ approaches very small values.
As shown in Fig.~\ref{fig3}(a), the decay exponents $\alpha$ for three different system sizes takes similar small values, indicating a delocalized to localized transition at $\Delta_C \sim 5.0 \pm 0.3$. 
The  charge imbalance $I(t)$ almost saturates to a stationary  value near $\Delta_C \gtrsim 5.0$ 
and does not decay with time [see inset of Fig.~\ref{fig3}(a)], also confirming the localization of charge.

Figure~\ref{fig3}(b) contains the $\alpha$ exponent for the disordered Hubbard model vs $\Delta$.
Compared to the disorder extended Hubbard model (at $U=2V=4$) the doublons are more stable 
at $U=4$ (see Fig.~\ref{fig8}), and
for this reason the product state with doublons are less delocalized. 
The decay exponent takes comparatively smaller values  even in the delocalized phase 
[see Fig.~\ref{fig3}(a) and Fig.~\ref{fig3}(b)].
As shown in Fig.~\ref{fig3}(b), the decay exponents $\alpha$ for three system sizes coincides 
for $\Delta \gtrsim 4.0$, and the charge imbalance $I(t)$ at $\Delta = 4.0$ almost saturated with 
time $t$ [see inset of Fig.~\ref{fig3}(b)], yielding as the estimate for the delocalized to localized 
transition at $\Delta \sim 4.0 \pm0.3$. Interestingly, 
the transition from the delocalized to localized phase occurs at smaller values of $\Delta$ for the
disordered Hubbard model (at $U=4$), when compared to the extended-Hubbard model (at $U=2V=4$). The 
critical disorder $\Delta_C$ for the Hubbard model decreases with increase in $U$, because 
with increase in on-site interaction $U$ the
doublons are much more stable [see Fig.\ref{fig8}(b)]. The doublons tunnel from one site to another only 
 through second-order hopping process, which leads to strong localization even for smaller 
values of disorder strength~\cite{mschre}.
However, for the extended Hubbard model at $U=2V$ a doublon can break into single fermions 
even for strong interactions [see Fig.\ref{fig8}(a)], which induces a larger $\Delta_C$.

In summary, the numerical analysis presented above [and also the 
long-time dynamics of $I(t)$ shown in Appendix Fig.~\ref{fig14}(a)] 
gives evidence that the product state with doublons behaves as a MBL state,
even in the presence of spin-SU(2) symmetry and strong interactions.
 The localization of charge and spin [see Fig.~\ref{fig9}] for larger values of $\Delta$, suggest that this product state has 
overlap with the non-thermal sector (area-law eigenstates) of the system Hamiltonian~\cite{ivan}.
In an interesting study on a spin-disordered Hubbard model with pseudo-spin SU(2) symmetry, 
 it has been argued that  the system has both a large number of area-law and log-law eigenstates~\cite{Xiong}. 
Furthermore, from the time evolution they demonstrate that the product state (quarter-filled singlons), 
 which has overlap with area-law eigenstates, shows all the dynamical properties of a MBL~\cite{Xiong}.  

\subsection{Quench from the product state of singlons}

In this subsection, we focus on the effect of the long-range interactions $V$ (with $V$ the 
prefactor of the dipolar-interaction $1/{r^3}$) on the 
localization of the spin degree of freedom when subject to a strong quasi-periodic potential.
Our primary aim is to investigate the localization of spin with a long-range spin-SU(2) symmetric Hamiltonian, 
in both the large and small $U$ interaction limits. As discussed in Sec.~I, it is believed at present that
with short-range SU(2)-symmetric models, spin localization does not occur. 
Including strong disorder, the charge degree of freedom freezes [see Fig.~\ref{fig13}(e)], 
but the spin in principle could delocalize by virtual hoping processes between different sites~\cite{ivan,trotzky}. 
However, the virtual hoping process depends strongly on the density of singly-occupied sites~\cite{ivan}.

Here, in order to study the dynamics of the spin degree of freedom, 
we started the time evolution using as initial state a N\'eel state $|\uparrow \  \downarrow \  
\uparrow \  \downarrow \  \uparrow \  \downarrow \  \uparrow \  \downarrow \ \uparrow \  \downarrow ..\rangle$
(where all sites are singly occupied with spin up ($\uparrow$) and spin down ($\downarrow$) on alternative sites, 
allowing for the possibility of fast spin dynamics). 
To probe the spin localization, we use the local spin imbalance~\cite{fiemini} 
$I_S(i)=s^z_i-s^z_{i+1}$, where $s_i^z=\frac{1}{2}\left(n_{i \uparrow}-n_{i\downarrow}\right)$ is the local spin operator
at site $i$.

\subsubsection{Large interaction $U$ limit}
\begin{figure}[h]
\centering
\rotatebox{0}{\includegraphics*[width=\linewidth]{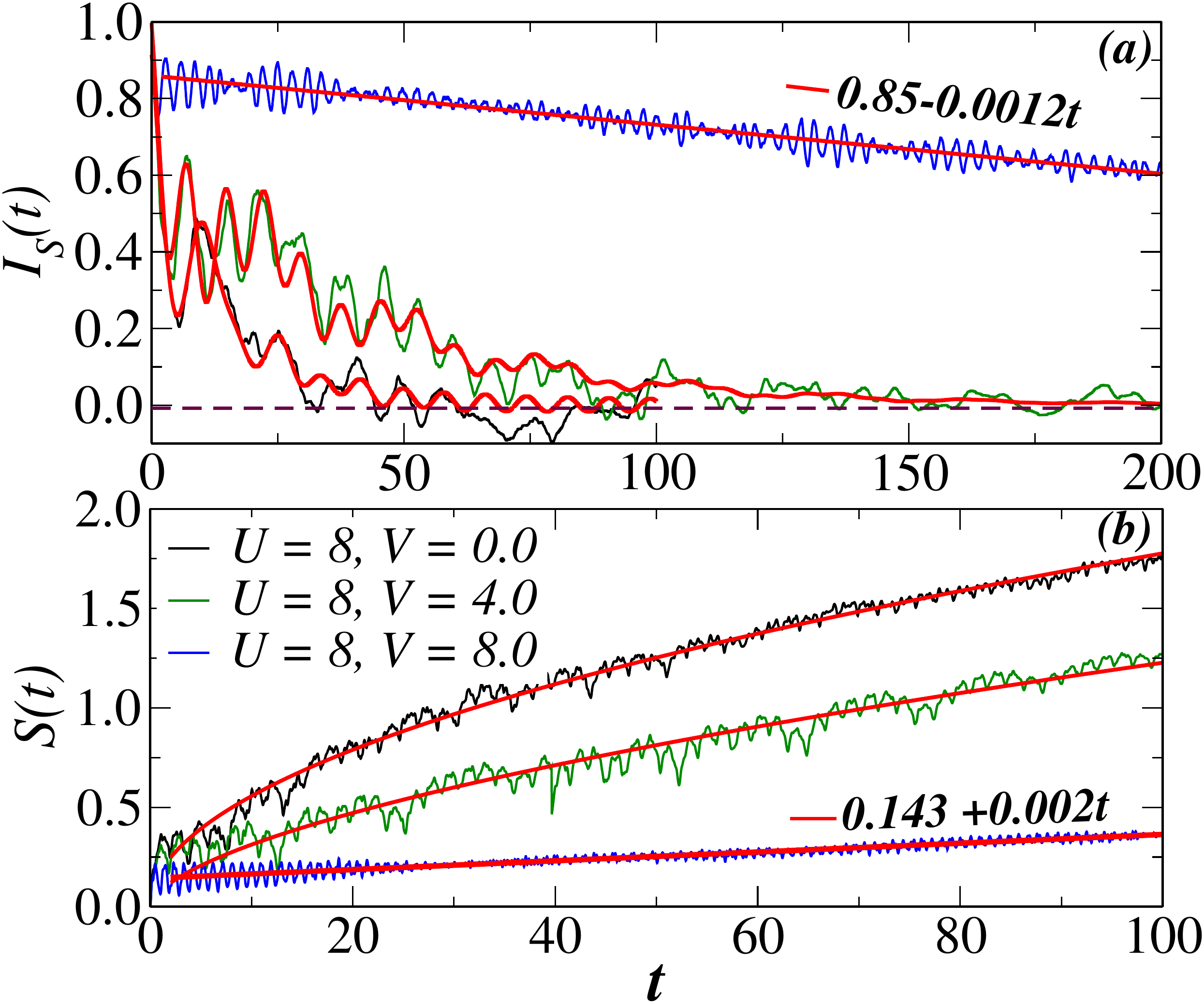}}
\caption{(a) Time evolution of the spin imbalance $I_S(t)$, with $t$ in units of $\hbar/t_{hop}$. 
 The red curve represents the fit to $I_S(t)$ by using the fitting functions described in the text.
(b) Entanglement entropy $S(t)$ vs time $t$. The red curve contains the power-law fit to $S(t)$.
These results are obtained by ED at $\Delta=16$, using system sizes $L=12$ (for $I_S(t)$) and $L=10$ (for $S(t)$),
 in the large interaction $U$ limit.} 
\label{fig4}
\end{figure}
We start the analysis of spin localization with a strong on-site interaction $U=8$, performing 
the time evolution using ED for a system
size $L=12$ and  disorder strength $\Delta =16$.  Figure~\ref{fig4}(a) shows 
the averaged spin imbalance $I_S(t)=\frac{2}{L} \sum_i (-1)^i s^z_i$, 
i.e. the average of the spin over all sites, followed by  averaging over six different disorder phases $\phi$.
For $U=8$ and $V=0$ (the standard on-site Hubbard model with disorder), the spin imbalance $I_S(t)$ decays exponentially and
 oscillates around zero after $t \gtrsim 50 \hbar/t_{hop}$. 
The very fast decay of the spin imbalance is due to the strong interaction and to the single occupancy of fermions (with spin up and spin down  on alternative sites) 
 of the N\'eel state, which enhances the exchange processes and leads to a faster spin delocalization.
For $U=2V=8$, the spin imbalance $I_S(t)$ still decays exponentially,
but comparatively at a slower pace than at the ($U=8$, $V=0$) case, and approaches zero after $t \gtrsim 150 \hbar/t_{hop}$.
Based on the studies of spin dynamics in Heisenberg models~\cite{punk1,punk2,luitz},
to describe the time evolution of spin imbalance  $I_S(t)$,  we have used the fitting function
$I_S(t)=a e^{-t/\tau} cos\left(\omega_1 t +\theta\right) + b e^{-\eta t}+c e^{-\zeta t} sin(\omega_2 t+ \theta)$.
The first term of the fitting function captures the fast exponentially decaying oscillations
and it is similar to the clean case of the anisotropic Heisenberg model~\cite{punk2}. The second term describes the  
exponential decay of spin imbalance with exponent $\eta$, while the third term contains
the characteristic oscillations for spin imbalance $I_S(t)$~\cite{luitz}.
We found that the  decays of spin imbalance for  both the case $U=8$ at $V=0$ and the case $U=2V=8$ are approximately
 described by the exponential fitting function, with decay exponents $\eta=0.073$ and $0.023$, respectively. Thus,
in the regime analyzed thus far, spin delocalization occurs.

Interestingly, at $U=V=8$, i.e. increasing $V$, the  spin imbalance shows a drastically different behavior. 
In this case the decay of spin imbalance $I_S(t)$ is reduced strongly and now it can be phenomenologically described using a linear fitting function 
$I_S(t)= m t+c$, with a very small decay rate $m=-0.0012$. This is qualitatively different from the previous 
exponential decay.
Based on second-order perturbation theory and using just two sites,
we find that the nearest-neighbor term $V$ causes the decrease in exchange processes, 
which leads to a strong reduction of spin delocalization
at least for the finite time scales explored here (see Appendix for more details).

The SU(2) symmetry of the Hamiltonian changes the entanglement 
structure of the eigenstates~\cite{protop,Xiong}, which leads to the faster growth of entanglement with time after a quantum quench.
The previous studies based on the disordered short-range Hubbard model with spin-SU(2) symmetry, find that the time evolution of entanglement
entropy grows as a power-law with time (instead of logarithmic growth), even in the presence of strong disorder~\cite{prelov}.
Here in order to study the behavior of entanglement entropy $S(t)$, with long-range interaction and spin-SU(2) symmetry, 
for the product state with singlons, in Fig~\ref{fig4}(b) we have plotted $S(t)$,
averaged over six different disorder phases $\phi$, for a system size $L=10$.
At $U=8$ and with $V=0$ the entanglement entropy $S(t)$ exhibits power-law 
growth with time $t$ ($S(t) \sim 0.17 t^{0.50}$), which is similar to previous results of SU(2) symmetric on-site disordered Hubbard model.

Interestingly, at $U=V=8$, the entanglement entropy changes qualitatively its behavior.
 Now it increases linearly with time rather than via a sub-linear power law. More importantly, 
the growth is very slow following $S(t)=0.143+0.002t$ with a small slope $m=0.002$. 
 The slow linear decrease in the spin imbalance previously described, and the slow linear growth of the entanglement entropy with a small slope shown here at $U=V$, 
provide evidence of a drastic reduction of spin-delocalization as compared to the on-site Hubbard model (at $U=8$ with $V=0$).
Surprisingly, we found that the phenomena of slowdown of spin delocalization at $U=V=8$ does persist
also for other values of $U=V$ (see  Fig.~\ref{fig12} in Appendix).

\subsubsection{Small interaction $U$  limit}
\begin{figure}[h]
\centering
\rotatebox{0}{\includegraphics*[width=\linewidth]{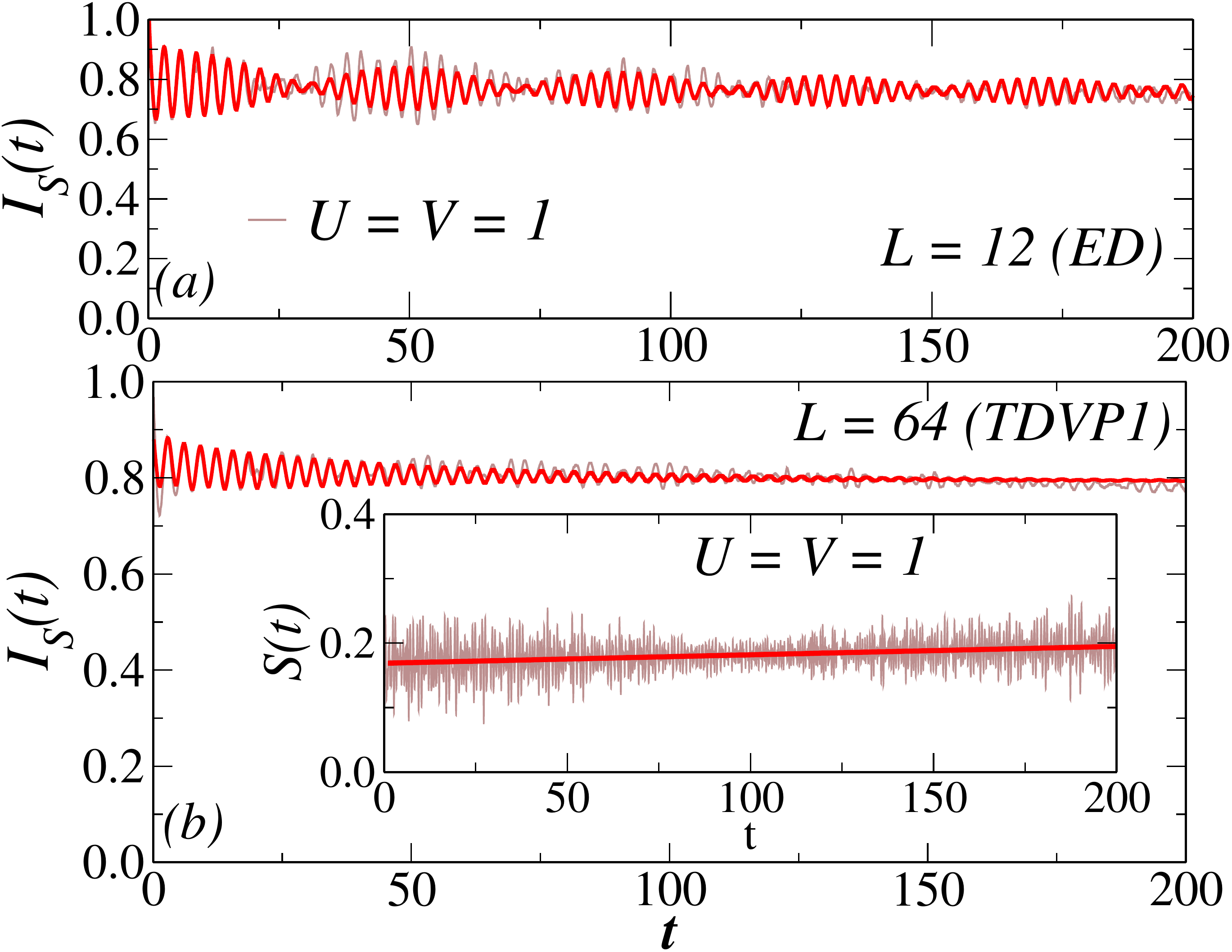}}
\caption{ Time evolution of the spin imbalance $I_S(t)$, with  $t$ in units of $\hbar/t_{hop}$,
in the small interaction limit $U=V=1$ and with strong disorder $\Delta=12$.
(a) Results using ED for a system size $L=12$. The red curve represents the power-law fit to $I_S(t)$ using the 
fitting function described in the text.
(b) Results using TDVP1 at $\Delta=12$ for a system size $L=64$. The inset shows the time evolution of the 
entanglement entropy $S(t)$ vs time $t$. 
The red curve displays the linear fit to $S(t)$.}
\label{fig5}
\end{figure}

Furthermore, to better visualize the role of interactions on spin dynamics,
here we study the dynamics of the spin degree of freedom in the small interaction $U$ limit. 
We consider on-site interaction $U=1$ and disorder strength $\Delta=12$.
 Figure~\ref{fig5}(a) shows the average spin imbalance $I_S(t)$ for a system size $L=12$ at $U=V=1$. 
The decay of $I_S(t)$ can be described by using the fitting function 
$I_S(t)=a e^{-t/\tau} cos\left(\omega_1 t +\theta\right) + b t^{-\eta }+c t^{-\zeta} sin(\omega_2 t+ \theta)$,
as suggested in Ref.\cite{luitz} for the disordered Heisenberg model. 
 The spin imbalance decay term, with a small exponent $\eta=0.010$, 
implies the very slow delocalization of the spin degree of freedom with time. 

To confirm this effective localization of spin using larger system sizes, 
in  Fig.~\ref{fig5}(b) we show the spin imbalance $I_S(t)$ vs $t$  for a system size $L=64$, using the TDVP1 time-evolution method. 
For this TDVP1 time evolution, we approximate the long-range interaction $V$ by only a nearest-neighbor 
interaction $V=1$~\cite{footnote}. 
 Using $L=64$, the spin imbalance can still be described by the same fitting function 
$I_S(t)=a e^{-t/\tau} cos\left(\omega_1 t +\theta\right) + b t^{-\eta }+c t^{-\zeta} sin(\omega_2 t+ \theta)$ 
 with  exponent $\eta=0.011$, very close to the 0.010 of the smaller system. Finite size effects appear
to be small for $I_S(t)$. 
The abnormally small value $\eta=0.011$ suggests localization of spin in the weak coupling limit 
(at least for the time scales of our numerical simulations).
Interestingly, at  $U=V=1$ we find a very slow linear increase in the entanglement entropy,
which can be described using $S(t)=0.16+0.0001t$, with a very small slope $m=0.0001$ [inset of  Fig.~\ref{fig5}(b)].
Compared to the strong $U$ limit, we find that the slowdown of spin delocalization is more effective at small $U$, 
i.e. the spin degree of freedom is more localized at weak $U$ coupling.
\begin{figure}[h]
\centering
\rotatebox{0}{\includegraphics*[width=\linewidth]{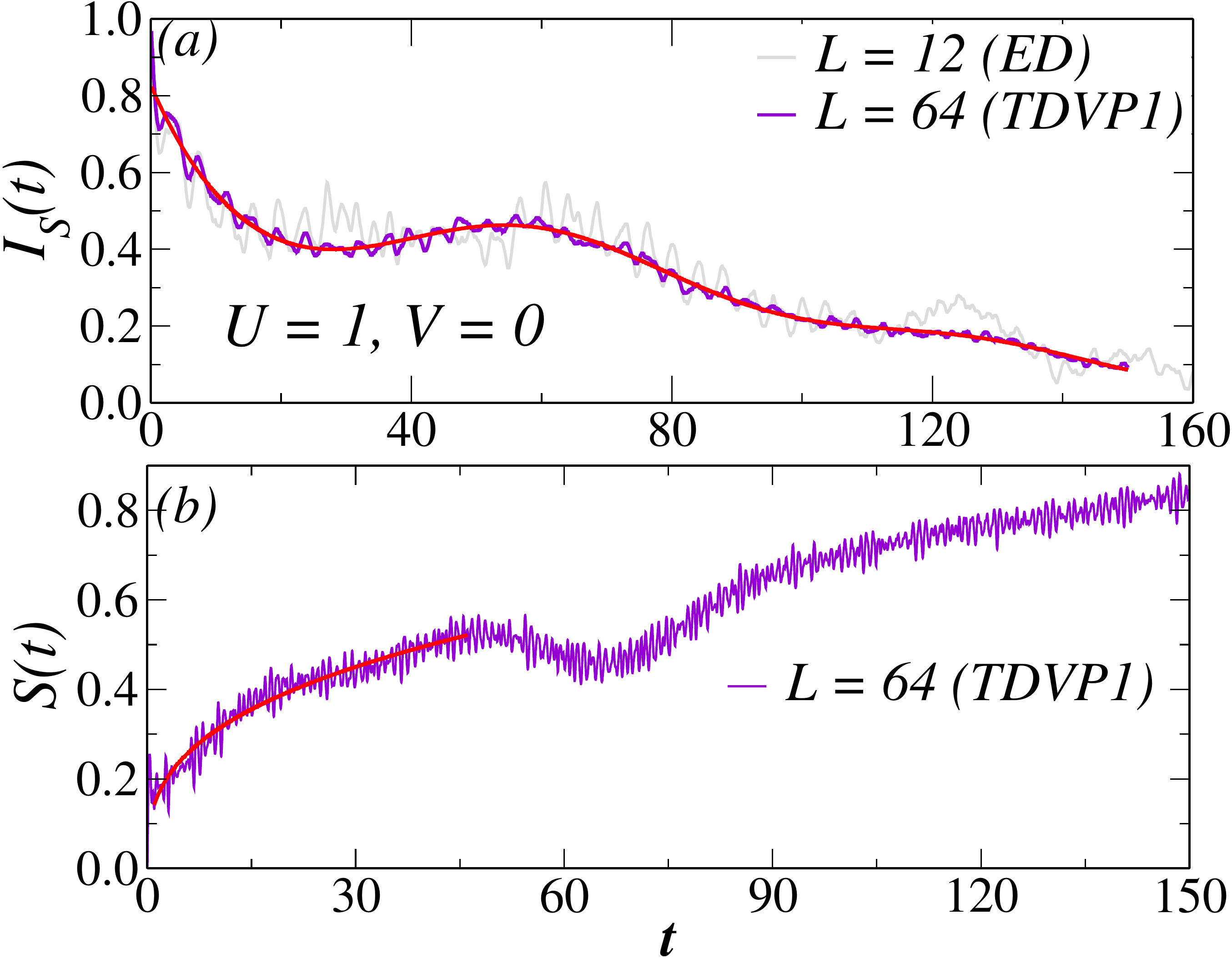}}
\caption{(a) Time evolution of the spin imbalance $I_S(t)$, with $t$ in units of $\hbar/t_{hop}$,
for the disordered on-site Hubbard model at small $U=1$ and $V=0$. 
 The red curve represents the power-law fit to $I_S(t)$, using the fitting function described in the text.
(b) Entanglement entropy $S(t)$ vs $t$. The red curve shows the power-law fit to $S(t)$.
These result were obtained using TDVP1 at $\Delta=12$ and for system size $L=64$.} 
\label{fig6}
\end{figure}

 Figure~\ref{fig6}(a) contains the average spin imbalance $I_S(t)$ vs time $t$, 
for the disordered on-site Hubbard model at $U=1$ and $V=0$, using two different system sizes $L=12$ (ED)
and $L=64$ (MPS). The decay of the spin imbalance $I_S(t)$ for both sizes $L=12$ and $L=64$ follows a similar behavior 
(except for some larger oscillations in the $L=12$ case). This suggests
that finite-size effects are very small for the local observable $I_S(t)$. 
The spin imbalance $I_S(t)$ decays to approximately zero at $t\gtrsim 160 \hbar/t_{hop}$,
showing delocalization of the spin degree of freedom. 
The overall decay of the spin imbalance $I_S(t)$ can be well described by a fitting function
$I_S(t)=a e^{-t/\tau} cos\left(\omega_1 t +\theta\right) + b e^{-\eta t}+c e^{-\zeta t} sin(\omega_2 t+ \theta)$, with exponential decay exponent $\eta=0.019$.
Interestingly, starting the time evolution with the N\'eel state, we find that the delocalization of the spin degree of freedom becomes quite faster ($U=1,V=0$),
as compared to previous studies based on the on-site 
disordered Hubbard model with a different initial product state~\cite{prelov}.
The entanglement entropy $S(t)$ at $U=1$ also follows a similar pattern as the spin imbalance $I_S(t)$,
and grows algebraically with much faster rate compared to the extended Hubbard model [at $U=V=1$ and $\Delta=12$]. 
The $S(t)$ initially grows as a power-law with $S(t)=0.14t^{0.34}$ for $t<50 \hbar/t_{hop}$. 
After a small decrease in S(t) [for the same time window where $I_S(t)$ increases],
 at longer times $S(t)$ grows at a faster rate [see  Fig.~\ref{fig6}(b)].

Summarizing, the inclusion of a long-range SU(2) symmetric term in the Hamiltonian
drastically slows down the delocalization of the spin degree of freedom, and qualitatively 
alters the time dependence when compared with short-range SU(2) symmetric models. 
We also find that in the strong $U$ limit the delocalization of spin is faster compared to the small $U$ limit [see Fig.~\ref{fig4}(a) and Fig.~\ref{fig6}(a)].

\subsection{Localization of charge and spin in full energy spectrum}
\begin{figure}[h]
\centering
\rotatebox{0}{\includegraphics*[width=\linewidth]{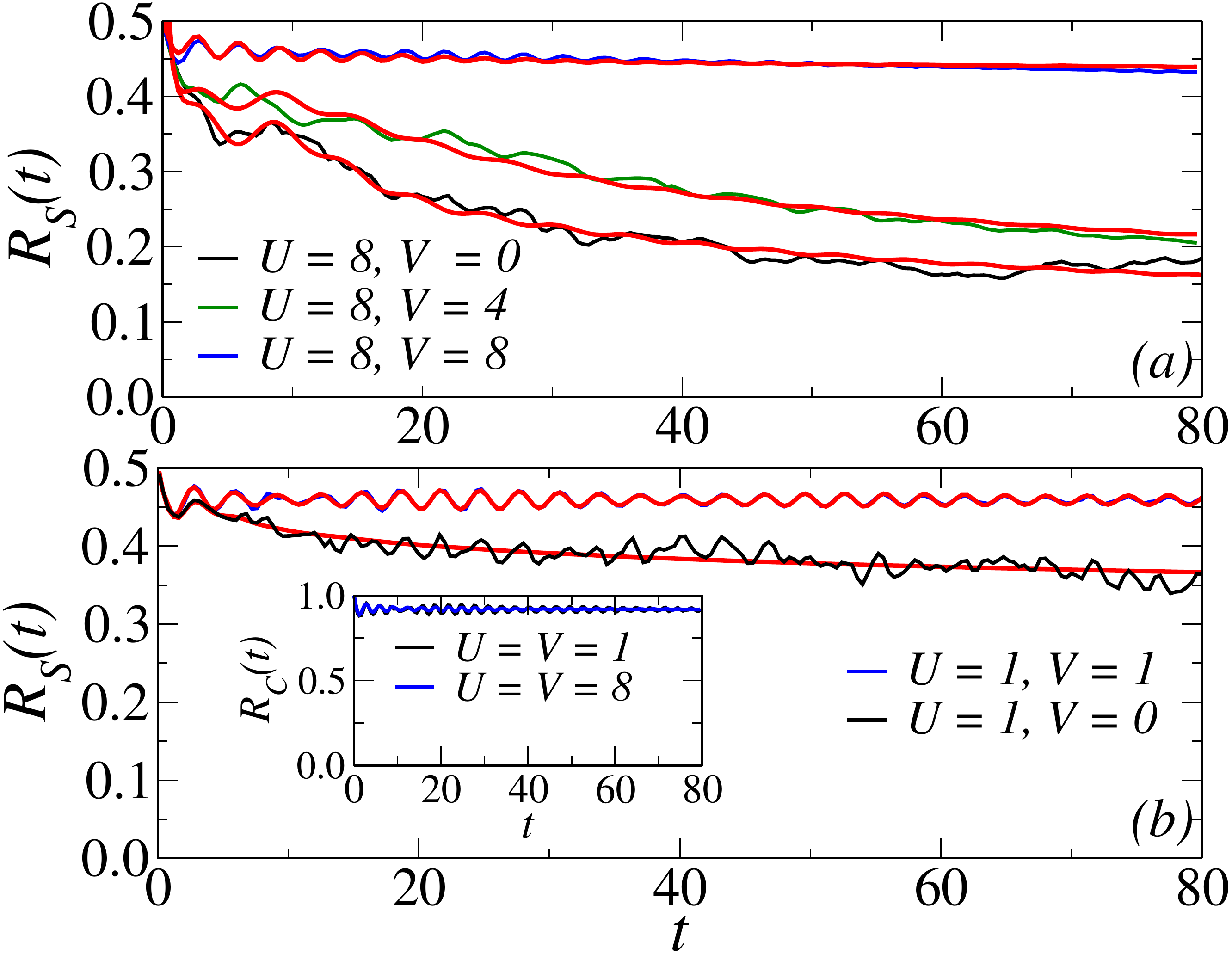}}
\caption{Averaged spin auto-correlation $R_S(t)$ vs time
$t$ (in units of $\hbar/t_{hop}$) for the cases of (a) large interaction limit $U=8$ and (b) small interaction limit $U=1$.
The red curves represent power-law fittings of $R_S(t)$  with exponent $\eta$.
The inset shows the charge auto-correlation $R_C(t)$ vs $t$.
These results were obtained using the full-diagonalization method of a system size $L=8$  
 at $\Delta=16$.} 
\label{fig7}
\end{figure}

Thus far, we have discussed the non-equilibrium dynamics for a 
quantum quench starting with two extreme product states at half-filling, 
one with doublons and another with singlons. Such a procedure is quite useful in cold-atom experiments. For a fully many-body localized MBL system, the charge and spin both should be localized in the entire many-body spectrum of the system Hamiltonian~\cite{papic,ogan}.
As discussed in previous studies~\cite{panda,protop,potter}, a full many-body localization is not possible with a continuous non-Abelian symmetric system, due to the appearance of degeneracies in the energy spectrum. 
Here, to study the extent of localization of charge and spin in the full energy-spectrum in the presence of a long-range interaction $V$,
 we calculated the local charge and spin auto-correlation~\cite{modak,nag}
by full diagonalization of the system Hamiltonian Eq.(1) on a chain with $L=8$ sites and at large values of
disorder $\Delta=16$. The spin auto-correlation function is defined as
   
\begin{eqnarray}
R_S(i,t)=\frac{4}{D}\sum_n \langle \psi_n| s^z_i s^z_i(t)|\psi_n\rangle \nonumber \\
 = \frac{4}{D} \sum_{m,n}\exp{\big(-it(E_n-E_m)\big)}|\langle \psi_n|s^z_i|\psi_m\rangle|^2
\end{eqnarray} 
where $D$ is the Hilbert space dimension, $|\psi_m\rangle$ is the $m$-th eigenstate of the Hamiltonian, and $s^z_i$
is the local spin operator at site $i$.
 We explore the extent of spin localization, at large and small $U$, by measuring the spin auto-correlation 
averaged over all sites i.e, $R_S(t)=(1/L) \sum_i R_S(i,t)$, followed by
averaging over five different values of the phase $\phi$.
Similarly for the charge, we define the charge auto-correlation as $R_C(t)=\frac{2}{DL} \sum_i\sum_{n} \langle \psi_n| \tilde{n}_i \tilde{n}_i(t)|\psi_n\rangle$.
For a fully-localized system, both $R_C(t)$ and $R_S(t)$ are expected to saturate with increasing time to a value close to the initial value at $t=0$, 
namely remain finite even after a long time evolution. 
By contrast, a Hamiltonian with a finite fraction of delocalization would show a power-law
decay and the extent of delocalization  in the spectrum can be measured with an exponent $\eta$.
As shown in the inset of Fig.~\ref{fig7}(b), the charge correlation $R_C(t)$ for $U=V=8$ and $U=V=1$
remain constant with time, providing evidence of localization of the charge degree of freedom at $\Delta=16$. 
Figure~\ref{fig7}(a) contains the  spin-spin auto-correlation for the SU(2) spin-symmetric Hamiltonian Eq.(1),
at large $U$. We observed that the power-law decay of  $R_S(t)$ can be captured by the fitting function 
$R_S(t)=a e^{-t/\tau} cos\left(\omega_1 t +\theta\right) + b t^{-\eta }+c t^{-\zeta} sin(\omega_2 t+ \theta)$.
In the case of the long-range Hubbard model (with $U=V=8$), the spin-correlation  
$R_S(t)$ decays with a small exponent $\eta=0.017$, hinting to localization of the spin degree of freedom 
in a large fraction of the energy spectrum.
On the other hand at  $U=2V=8$ and at $U=8, V=0$,
$R_S(t)$ decays with a larger exponent $\eta = 0.335$  and $\eta = 0.344$, respectively, 
indicating delocalization of the spin degree of freedom over a large fraction of the energy spectrum. 
 
Figure~\ref{fig7}(b) displays the spin-correlations $R_S(t)$ at small $U$.
For the long-range Hubbard model ($U=V=1$), $R_S(t)$ remains constant with time  ($\eta=0.0003$, i.e. zero within our accuracy).
Thus, the spin auto-correlation  [in Fig.~\ref{fig7}(b)] and the charge auto-correlation at $U=V=1$
[inset of  Fig.~\ref{fig7}(b)] suggest localization of spin and charge in the full energy spectrum
(at least for a finite system and for a finite time). 
On the other hand, for the disordered on-site Hubbard model there is  delocalization of the spin
degree of freedom, with $R_S(t)$ decaying as a power law with a larger exponent $\eta \sim0.066 >> 0.0003$.    
The restoration of a full-MBL using an on-site interacting disordered Hubbard model has been shown in a recent study~\cite{sroda} 
by breaking the SU(2) symmetry of the Hamiltonian using asymmetrical hopping $t_{\uparrow} \neq t_{\downarrow}$. 
Here we show that localization of charge and spin in the full energy spectrum (at least for the time scale of our simulation) can occur
 without breaking the SU(2) symmetry but instead simply rendering the interactions long-range instead of short-range.

\section{Conclusions}
In conclusion, we have explored the localization  
of charge and spin degrees of freedom for fermions with long-range interactions,
in the presence of a quasiperiodic potential to add disorder and for experimentally relevant time scales. 
Our Hamiltonian has a spin-SU(2) symmetry. 
We studied the dynamics of charge and spin using two different initial states.
Specifically, we considered two extreme cases of product states: 
 doublons  $|\uparrow \downarrow 0 \uparrow \downarrow 0 
\uparrow \downarrow 0 \uparrow \downarrow 0 \uparrow \downarrow \rangle$ and singlons $|\uparrow \  \downarrow \  \uparrow \  \downarrow  
 \ \uparrow  \ \downarrow \  \uparrow  \ \downarrow \  \uparrow \ \downarrow \rangle $,  at half filling.
Compared to the on-site (short-range) Hubbard model, in the long-range Hubbard model at $U \sim 2V$ 
we show that at large interaction $U$ and without 
disorder the doublons are found to be unstable. This leads to 
larger  values of $\Delta_C$ for the delocalized to localized transition in the case of the extended Hubbard model.
In fact, for the product state with doublons, at strong interaction $U=2V=8$, we observed a transition from delocalized 
to localized phases varying $\Delta$, by estimating the relaxation exponent of the charge imbalance.
Moreover, we found that for large disorder, the transport of charge and spin stops, 
whereas the entanglement entropy increases logarithmicaly with time.
Then, the product state with doublons acts as a MBL state (irrespective of the spin-SU(2) symmetry).
Starting the time evolution from the product state with doublons 
all of the dynamical properties of the MBL described above should be observed experimentally in dipolar fermionic systems. 
However, the system is not a full-MBL at $U=2V$ because the spin degree of freedom
delocalizes over a large fraction of the energy spectrum [see Fig.~\ref{fig7}(a)].

Starting the time evolution with the product state of singlons (N\'eel state),
we presented a comparative study of the disordered on-site Hubbard and disordered long-range Hubbard models
both at large and small interaction $U$ limits.   
In the large interaction $U$ limit but without long-range, i.e. $V=0$, the spin imbalance decays 
exponentially, compared to algebraic decay of spin for on-site disordered Hubbard model in previous publications. 
However, in contrast to previous studies with short-range
Hubbard models we found that including a long-range interaction in the Hamiltonian (i.e. $V$ nonzero), 
the decay of spin-imbalance slows down dramatically. This occurs {\it even without breaking the spin-SU(2) symmetry}. 
Eventually, at small $U$ when the long-range  $V$ and on-site $U$ interactions are equal in strength, the spin degree of freedom  localizes
(at least for the time scales of our numerical simulations).
Using the TDVP1 time-evolution method, we demonstrate that at small $U$ and keeping $U=V$, 
the localization of spin persists for the significantly larger systems typically used in cold atom experiments. 
In contrast to the algebraic growth of entanglement entropy in the disorder on-site Hubbard model,
 the entanglement entropy shows a slow linear growth for the extended-Hubbard model. 
Interestingly, the phenomena of slowdown of spin delocalization of spin also persist in the full energy spectrum
with the long-range interacting Hamiltonian. We observed that for large disorder the charge and spin both localizes in the full energy 
spectrum when the on-site and long-range interactions are equal in strength
[at least for a finite system and a finite time]. 

We have carried out the dynamics for long enough time and system sizes to reach 
experimentally relevant time scales and system sizes in which the localization of 
both charge and spin should be observed in dipolar systems subject to a quasiperiodic potential. 
However, due to the non-Abelian spin SU(2) symmetry there is the possibility that spin can delocalized
for much longer time scales.
Our work based on SU(2) symmetric long-range fermionic dipolar systems and using simple 
initial product states opens the possibility to  understand special many-body nearly localized systems 
where the relaxation rate is anomalously very slow and the system is expected to thermalize on much longer time scales\cite{Gopala}.

\section{Acknowledgments}
\begin{figure}[h]
\centering
\rotatebox{0}{\includegraphics*[width=\linewidth]{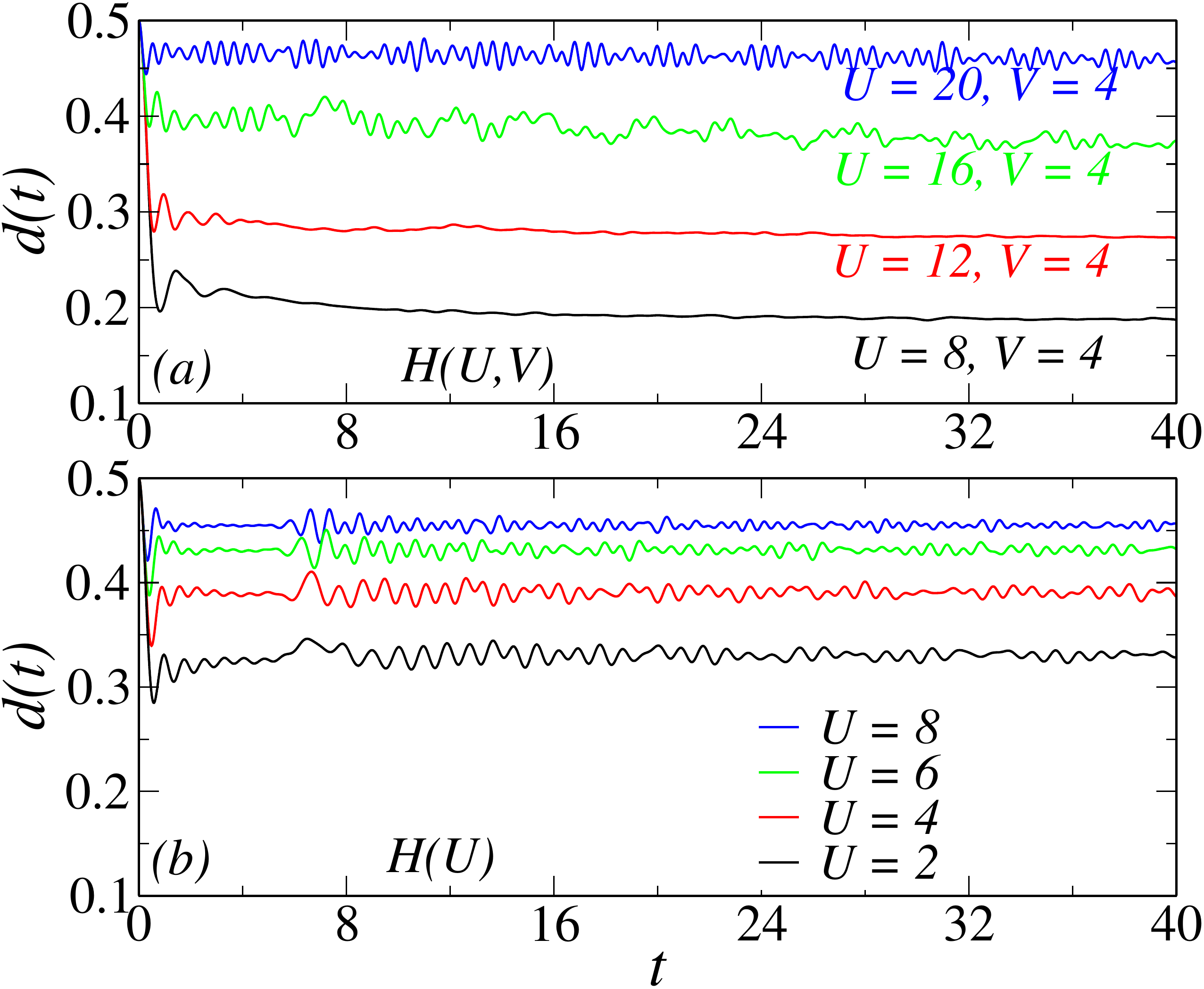}}
\caption{Time evolution of the site average double occupancy $d(t)$, vs 
time $t$ (in units of $\hbar/t_{hop}$)(a) long-range Hubbard model
(b)Hubbard model, for different values of interaction parameters.
 In both cases $L=12$, and $\Delta=0$ (no disorder).}
\label{fig8}
\end{figure}

We thank G. Roux and E. Quinn for useful discussions.
The work of B.P and E.D. was supported by the U.S. Department of
Energy (DOE), Office of Science, Basic Energy Sciences
(BES), Materials Sciences and Engineering Division.
S.K.P. acknowledges the Department of Science and Technology (DST), 
Science and Engineering Research Board (SERB), JCBose Fellowship, and the Govt. of India for financial support. 

\section{Appendix}

\subsection{Stability of doublons without disorder}
For the on-site Hubbard model (i.e. without long-range interactions $V=0$) 
in the large interaction limit ($U>>t_{hop}$), 
doublons are quite stable and can not decay to single fermions [see Fig.~\ref{fig8}(b)] for the energy-conserving time 
evolution we are carrying out~\cite{winkler}. 
On the other hand, for the extended Hubbard model with $V$ nonzero at nearest-neighbors, the doublons become unstable at $U=2V$ 
and can decay to single fermions [see Fig.~\ref{fig8}(a)] even with full energy conservation~\cite{Fhof}. 

Here we investigate the stability  of doublons by starting the time evolution with a product state of doublons 
$|\uparrow \downarrow 0 \uparrow \downarrow 0 \uparrow \downarrow 0 \uparrow \downarrow 0 \uparrow \downarrow \rangle$, 
using a long-range interacting Hamiltonian without any disorder i.e. $\Delta =0$. We find that the site average double occupancy, defined as
$d(t)= \frac{1}{L}\sum_i\langle n_{i\uparrow } n_{i\downarrow}\rangle$, at $U=2V=8$ 
decays quickly to a small but finite value and after a few oscillations
(Fig.~\ref{fig8}(a)). This indicates that at $U \sim 2V$ the 
doublons are partially unstable and can decay to individual fermions. 
On the other hand, for large values of onsite interactions $U=20$ and $V=4$ the number of doublons almost remains the same as the 
initial value at $t=0$, showing doublons are quite stable in this regime (Fig.~\ref{fig8}).
In the case of the on-site Hubbard model ($V=0$) even without any disorder ($\Delta=0$), with increase in  onsite interaction $U$ 
the doublons are very stable [see Fig.~\ref{fig8}(b)] and do not break into single fermions~\cite{Fhof}. 
However, in the long-range Hubbard model at $U = 2V$,
doublons are unstable providing a window to study a delocalized to localized transition with disorder
even at large interactions.

\subsection{Bipartite charge and spin fluctuations for product state with doublons}

\begin{figure}[h]
\centering
\rotatebox{0}{\includegraphics*[width=\linewidth]{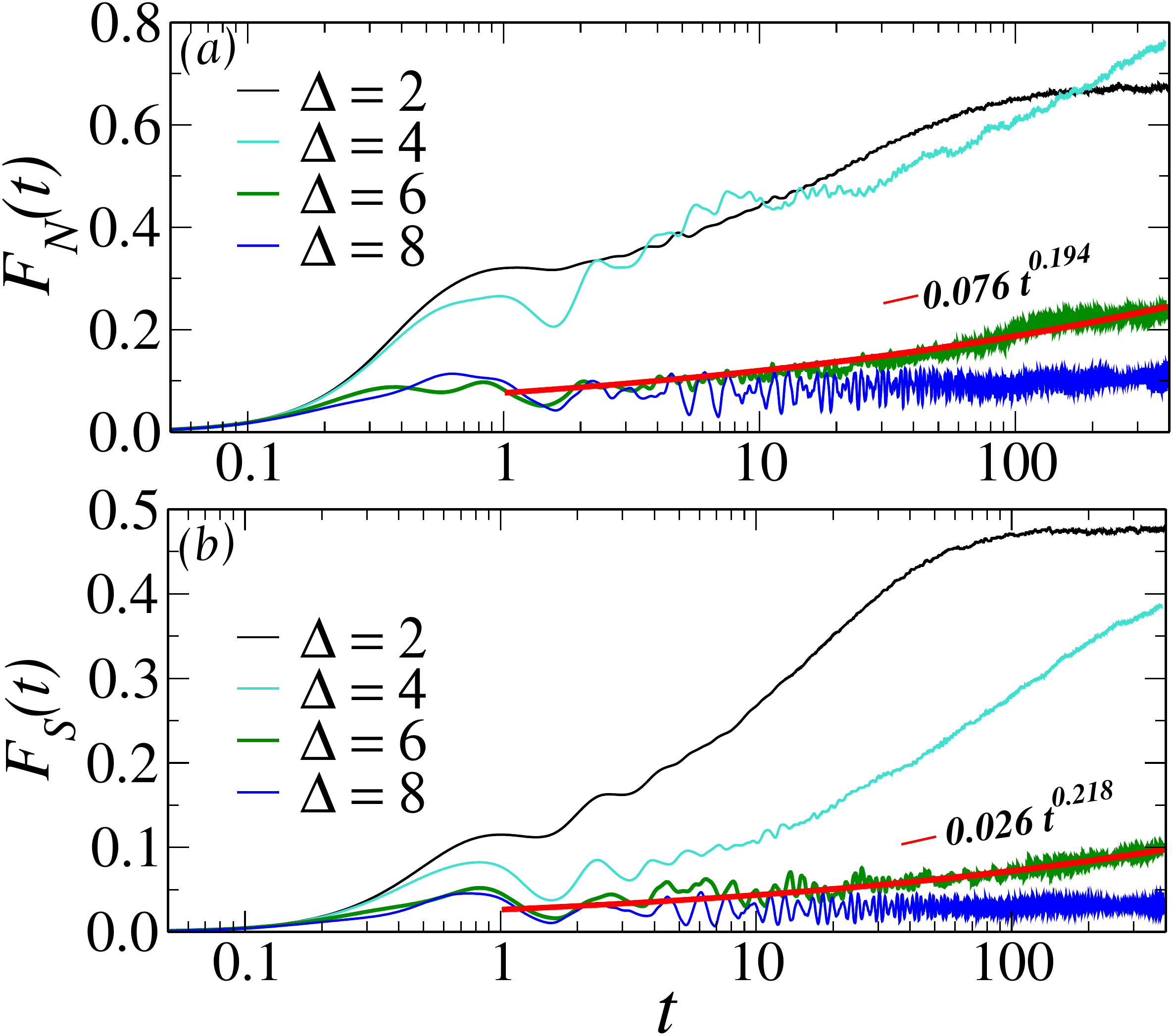}}
\caption{Time evolution of the bipartite fluctuations:  (a) charge-fluctuation $F_N(t)$ and (b) spin-fluctuation $F_S(t)$
 vs time $t$ (in log-scale and in units of $\hbar/t_{hop}$). We work at $U=2V=8$ and using
 several values of $\Delta$.
The red curves show a power-law fit to the charge and spin fluctuations near the transition point $\Delta=6$.} 
\label{fig9}
\end{figure}

To characterize the transport properties with increasing disorder strength $\Delta$ and  at interaction parameter $U=2V=8$,
we analyze the time-dependent bipartite charge and spin fluctuations [Fig.~\ref{fig9}(a) and Fig.~\ref{fig9}(b)], by dividing the system ($L=12$) in 
two equal parts: blocks A and B. The charge fluctuations of block A are defined as  
$F_N(t)=\langle \Psi(t)|N_A^2|\Psi(t)\rangle - \langle \Psi(t)|N_A |\Psi(t)\rangle^2$~\cite{hfran}, 
where $N_A=\sum_{i=1}^{L/2}\hat{n}_i$. 
Similarly, the $z$-component of the spin fluctuations of block A are defined as
$F_S(t)=\langle \Psi(t)| (S^z_A) ^2|\Psi(t)\rangle - \langle \Psi(t)|S^z_A |\Psi(t)\rangle^2$, 
where $S^z_A=\sum_{i=1}^{L/2}s^z_i$. These charge and spin 
fluctuations can be related to the transport of charges and 
$z$-component of spins from block A to B~\cite{maksym,rajeev}. 
For small values of disorder strength $\Delta$, the doublons can break into two 
single unpaired fermions (at $U=2V$), giving raise to a rapid enhancement in the 
charge $F_N(t)$ and spin $F_S(t)$ fluctuations in the system. 
Interestingly, near the phase boundary of the delocalized-localized transition 
($\Delta \sim 6.0$), the charge $F_N(t)$ and  spin $F_S(t)$ fluctuations
grow slowly as a power-law $\sim t^{\eta}$, with small exponent $\eta_N=0.194$ and $\eta_S=0.218$, respectively (see Fig.~\ref{fig9}). 
For large values of disorder  $\Delta = 8.0 $, the doublons are quite stable (they do not break into single fermions).
Doublons fluctuate close 
to the original lattice sites within some localization length. As a consequence, 
charge and spin fluctuations almost saturate to a small 
 value, indicating the suppression of transport of charge and spin degrees of freedom from block A to block B.    
Remarkably, the bipartite fluctuations can also be measured in ultra-cold atomic systems via quantum gas microscopy~\cite{mazurenko,alex,sherson}  
\subsection{ED versus TDVP time evolution}
Using the ITensor library~\cite{itensor}, we have implemented the one-site version of TDVP 
in finite matrix product states (MPS) for the time evolution of fermionic systems. Compared to other time evolution methods (t-DMRG or TEBD), the one-site TDVP preserves the energy of the system during the time evolution 
and does not depend on the Suzuki-Trotter decomposition of local terms of the system Hamiltonian~\cite{lubich,white}. 
In the TDVP method the time evolution of MPS is performed by
projecting the right hand side of the time-dependent Schr\"{o}dinger equation    
[$\frac{d|\psi(t)\rangle}{dt} = -iP_{MPS} H|\psi\rangle $] to the
tangent space of the variational MPS manifold of fixed bond dimension $\chi$~\cite{Cirac}. 
By this process, it constrains the time evolution of the system 
to a specific manifold of matrix product states with particular bond dimension $\chi$. 
In our work, first we have enlarged the bond dimension of the 
MPS, using the global subspace expansion (GSE) method~\cite{white} for the first few time steps of the time evolution. 
After reaching the desired value of bond dimension $\chi$, we have performed the time 
evolution with fixed bond dimension $\chi$ using one-site TDVP.
\begin{figure}[h]
\centering
\rotatebox{0}{\includegraphics*[width=\linewidth]{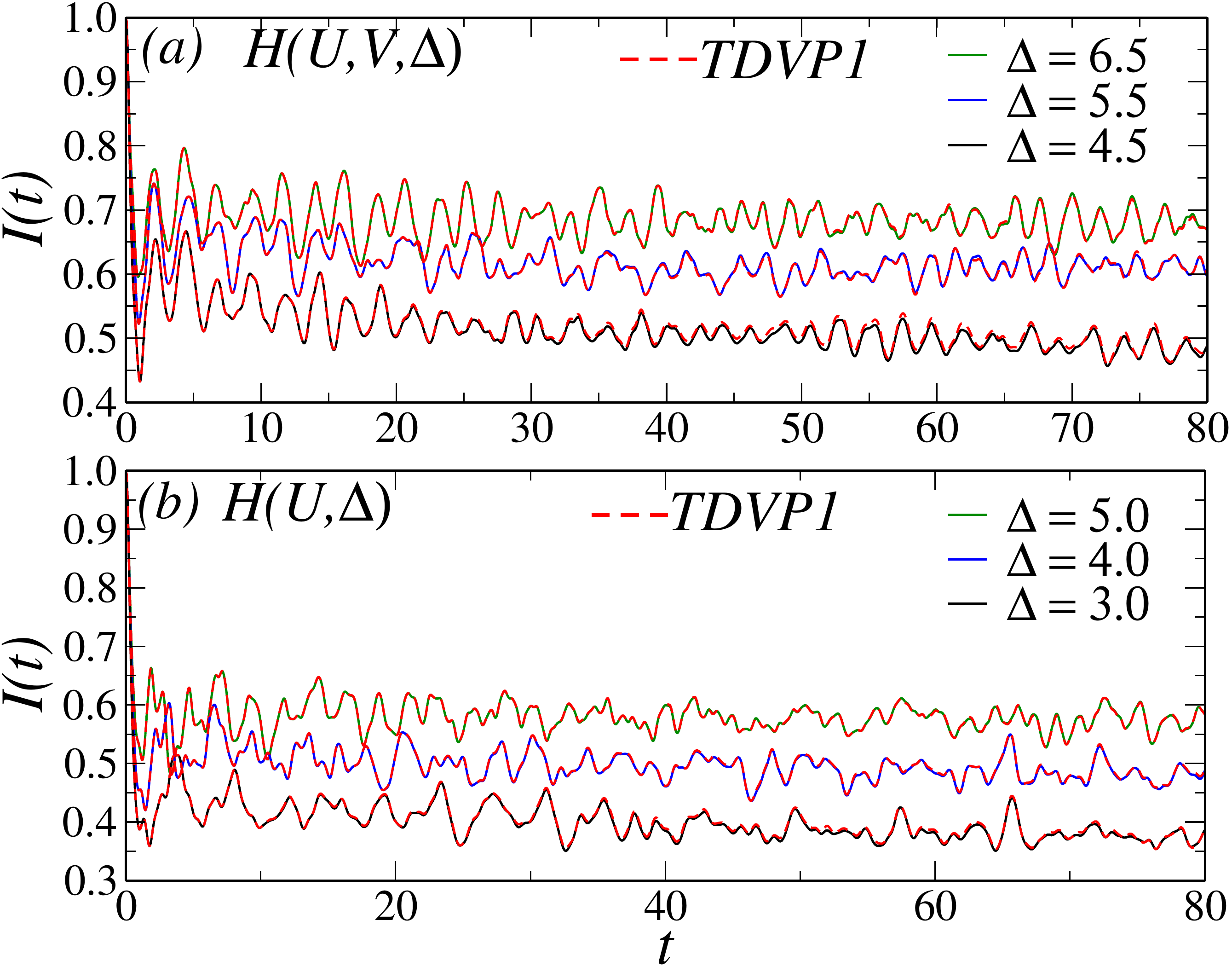}}
\caption{Comparison of the charge imbalance $I(t)$ near the transition region: using ED (solid lines) and TDVP(broken red lines) 
time evolution methods using system size $L=12$.
(a)For extended disordered Hubbard model at fixed value of $U=2V=4$ and for different values of $\Delta$.
(b)For disordered Hubbard model at $U=4$ and for different values of $\Delta$.
The charge imbalance $I(t)$ is averaged over sites followed by averaging over ten different values of phase factor $\phi$.}
\label{fig10}
\end{figure}

Figure~\ref{fig10} presents the comparison of the time evolution of charge imbalance $I(t)$,
 using the TDVP and ED methods for the disordered extended Hubbard [$H(U,V\Delta)$ $U=2V=4$]  and  on-site Hubbard [$H(U,\Delta)$ at $U=4$]
models for disorder values near the transition region. 
We have used time-step size $\delta=0.01$ (in units of $\hbar/t_{hop}$) and  10 to 12 time steps
to enlarge the bond dimension (using GSE-TDVP1) for the time evolution. We found that after 10 to 12 time steps
of the time evolution, the bond dimension of MPS reaches approximately $\chi=300$ to $500$ depending upon 
the system size, interaction, and disorder strengths. As shown in Fig.~\ref{fig10}(a), for the disorder value 
$\Delta=4.5$ 
(before the transition point), the $I(t)$ using TDVP shows a very small deviation from  ED results for large times, 
while for $\Delta \gtrsim 5.5$ ED and TDVP results 
matches quite accurately. On the other hand, for the disordered Hubbard model [Fig.~\ref{fig10}(b)]
the decay of $I(t)$ using TDVP follows the ED results near the transition region. These numerical results 
[Fig.~\ref{fig10}(a) and Fig.~\ref{fig10}(b)] shows that the TDVP method works quite well near the transition region involving the delocalized to localized transition.   

\subsection{Perturbative analysis for the N\'eel state}

At large  disorder strength $\Delta$, the charge degree of freedom localizes,
but the spin can interact through the exchange mechanism. For the disorder  on-site Hubbard model, 
in Ref.~\cite{ivan} by considering the virtual hopping between singly occupied distant sites, 
the effective spin Hamiltonian $H_{spin}=\sum_{r_i,r_j} J_{r_i,r_j} {\bf S}_{r_i} \cdot {\bf S}_{r_j}$ was derived.
The exchange coupling $J_{r_i,r_j}$  between two spins at distance $|r_i -r_j|$, has been obtained by using 
$2(r_i-r_j)$th order of perturbation expansion in the hopping amplitude. The initial states 
were the product state of singly occupied, or doubly occupied/empty sites~\cite{ivan}.

Here, in our case for the long-range interacting disordered Hubbard model 
using as initial state the N\'eel state $|\uparrow \  \downarrow \  \uparrow \  \downarrow  
 \ \uparrow  \ \downarrow \  \uparrow  \ \downarrow \  \uparrow \ \downarrow \rangle $, the most dominating exchange process occurs 
between nearest-neighbor sites (this exchange occurs from virtual hopping of a particle with spin $\uparrow$ to a neighboring site with spin $\downarrow$ or vice versa)~\cite{trotzky}. To gather insight into the physical picture related to the localization of the spin degree of freedom, we consider just two fermions interacting with  on-site $U$ and nearest-neighbor $V$ couplings,
at large disorder strength $\Delta >>t_{hop}$, and with the system initialized in a N\'eel state 
$|\uparrow , \downarrow \rangle$.
Starting with only two fermions $|\uparrow , \downarrow\rangle$ 
with opposite spins, we will use the canonical transformation of $H$:
\begin{equation}
\tilde{H}=\exp{\left(iS\lambda \right)} H \exp{\left(-i \lambda S \right)}.
\end{equation}
Here, $H=H_D+ \lambda H_{t_{hop}}$ in the basis $\big(|\uparrow, \downarrow \rangle, |\downarrow, \uparrow \rangle, |\uparrow \downarrow, 0 \rangle, |0, \uparrow \downarrow \rangle \big)$
where the diagonal part of the Hamiltonian matrix is
\[H_D= \left( \begin{array}{cccc}
\epsilon_1+\epsilon_2 & 0 & 0 & 0 \\
0 & \epsilon_1+\epsilon_2 & 0 & 0 \\
0 & 0 & 2\epsilon_1+U-V & 0 \\
0 & 0 & 0 & 2\epsilon_2+U-V \end{array} \right)\]
and the hopping term of the Hamiltonian matrix is
\[H_J= \left( \begin{array}{cccc}
0 & 0 & -t_{hop} & -t_{hop} \\
0 & 0 & t_{hop} & t_{hop} \\
-t_{hop} & t_{hop} & 0 & 0 \\
-t_{hop} & t_{hop} & 0 & 0 \end{array} \right).\]
To deduce the  hermitian matrix $S$, we used the condition $H_{t_{hop}}+i[S,H_D]=0$, which results
from requesting a vanishing first-order term in the hopping i.e.  
\[S= i\left( \begin{array}{cccc}
0 & 0 & -\frac{t_{hop}}{A} & -\frac{t_{hop}}{B} \\
0 & 0 & \frac{t_{hop}}{A} & \frac{t_{hop}}{B} \\
\frac{t_{hop}}{A} & -\frac{t_{hop}}{A} & 0 & 0 \\
\frac{t_{hop}}{B} & -\frac{t_{hop}}{B} & 0 & 0 \end{array} \right).\]
Here $A=(\epsilon_1-\epsilon_2)+(U-V)$ and $B=(\epsilon_2-\epsilon_1)+(U-V)$. After substituting 
the  $S$ matrix in the second-order term of the perturbative expansion in the hopping amplitude 
and projecting into the subspace  $\big(|\uparrow, \downarrow \rangle, |\downarrow, \uparrow \rangle\big)$,
we obtain the Hamiltonian $H_{spin} \sim J^{2nd}_{i,j} {\bf S}_i \cdot {\bf S}_j$ for two spins with exchange constant
\begin{equation}
J^{2nd}_{i,j}= -\frac{4t_{hop}^2 \big(U-V\big)}{\big(\epsilon_i -\epsilon_j \big)^2-\big(U-V\big)^2}. 
\end {equation}
We assumed $\Delta > 12$ and that there are no doublons formation during the time evolution. 
Here $\epsilon_i = \Delta \cos\big(2 \pi\beta i\big)$.
The form of the exchange $J^{2nd}_{i,j}$ allows us to explain qualitatively  
the behavior of the spin dynamics  when in presence of large disorder. 
For the N\'eel state  $|\uparrow \  \downarrow \  \uparrow \  \downarrow   \ \uparrow  \ \downarrow \  \uparrow  \ \downarrow \  \uparrow \ \downarrow \rangle $
 the most dominating exchange occurs from second order virtual hopping process~\cite{caveat-appendix}.
\begin{figure}[h]
\centering
\rotatebox{0}{\includegraphics*[width=\linewidth]{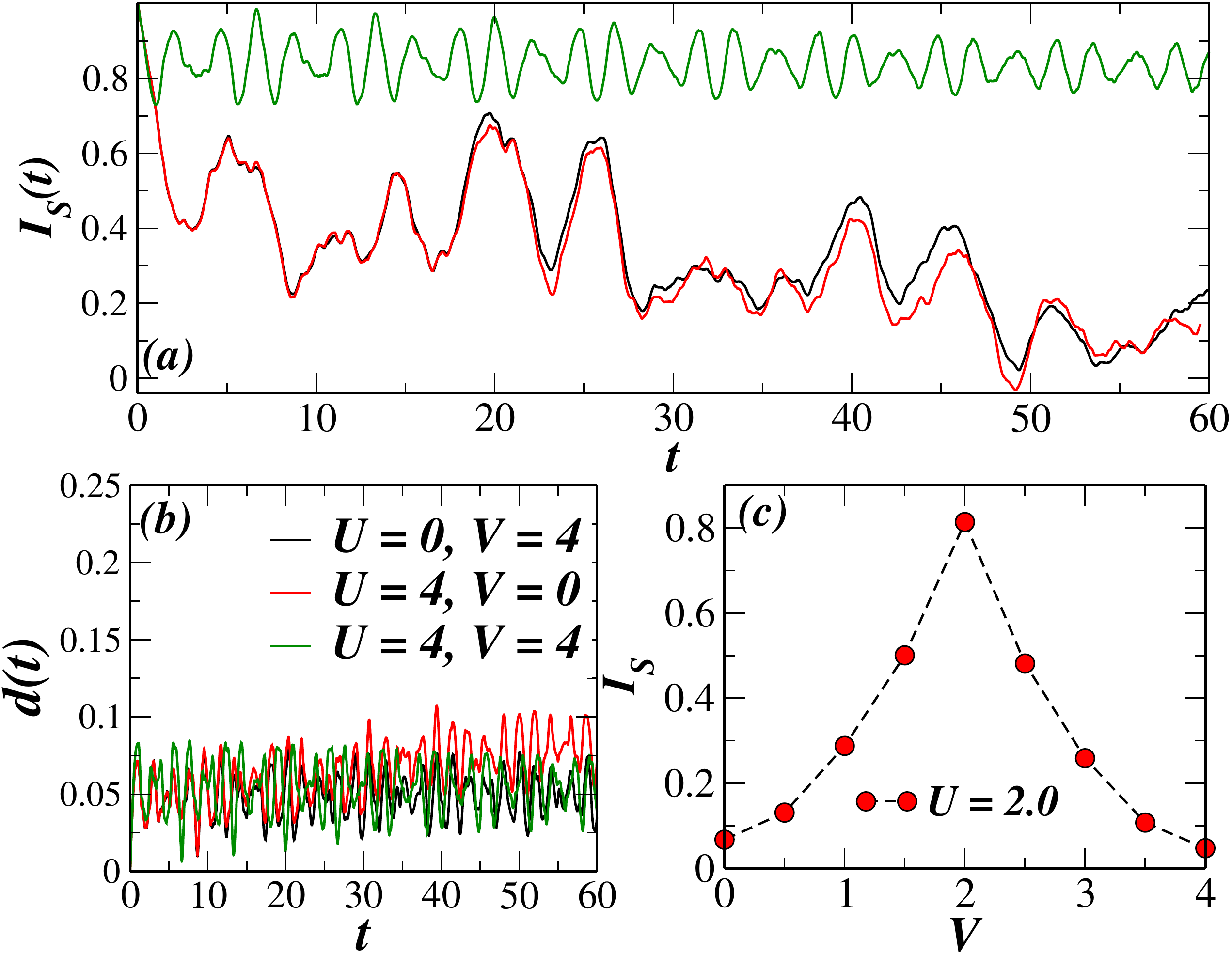}}
\caption{(a) Time evolution of the spin imbalance $I_S(t)$ vs time $t$ (in units of $\hbar/t_{hop}$)
for different interaction strengths. (b) Average doublons occupancy vs time $t$.
(c) Time-averaged spin imbalance $I_S$ vs $V$  at $U=2$,
showing that the spin imbalance acquires maximum values when $U=V=2$.  
These results are obtained using ED  at disorder strength $\Delta=16$, $\phi=0$, and for a system size $L=12$.} 
\label{fig11}
\end{figure}

	As shown in  Fig.~\ref{fig11}(a), for ($U=0$, $V=4$) and ($U=4$, $V=0$) the dynamics of spin imbalance $I_S(t)$ is quite similar 
up to intermediate times (the magnitude of $|J_{i,j}|$ is the same for these two values of interaction parameters). 
On the other hand, for $U=V=4$, the spin imbalance $I_S(t)$ remains close to initial values as the  exchange constant reduces to a smaller values 
providing evidence that the reduction in spin-delocalization of the $L=12$ sites fermionic system (at intermediate times) can be explained qualitatively
 by the second-order exchange process between two fermions. 
Figure~\ref{fig11}(b) displays the average doublons occupancy 
$d(t)=\sum_i \langle n_{i,\uparrow} n_{i,\downarrow}\rangle/L$ vs time $t$,
showing that for  $U=V=4$ there is no significant increase in the average number of doublons with time.
 Figure~\ref{fig11}(c) displays the time-averaged spin imbalance $I_S$ vs long-range interaction $V$, at $U=2$ and $\Delta=16$. 
For the time averaging of $I_S(t)$, we employ data from $t=150$ to $200$ in $\hbar/t_{hop}$ units.
With increase in $V$, the exchange process between fermions reduces (for $V\le U=2$), which leads to a decrease in spin
delocalization. As a result the average value of the spin imbalance $I_S$ increases with $V$.
On the other hand for $V>U=2$, the exchange process again increases by increasing $V$,
which leads to increase in spin delocalization. As a result $I_S$ decreases with increase in $V$, Fig.~\ref{fig11}(c). 
Interestingly, when on-site and long-range interactions are equal in strength
 the $I_S$ takes maximum values at $U=V=2$, showing the strong reduction in spin delocalization at $U=V$.

\subsection{Effects of interactions and disorder on spin and charge dynamics for N\'eel state}
\begin{figure}[h]
\centering
\rotatebox{0}{\includegraphics*[width=\linewidth]{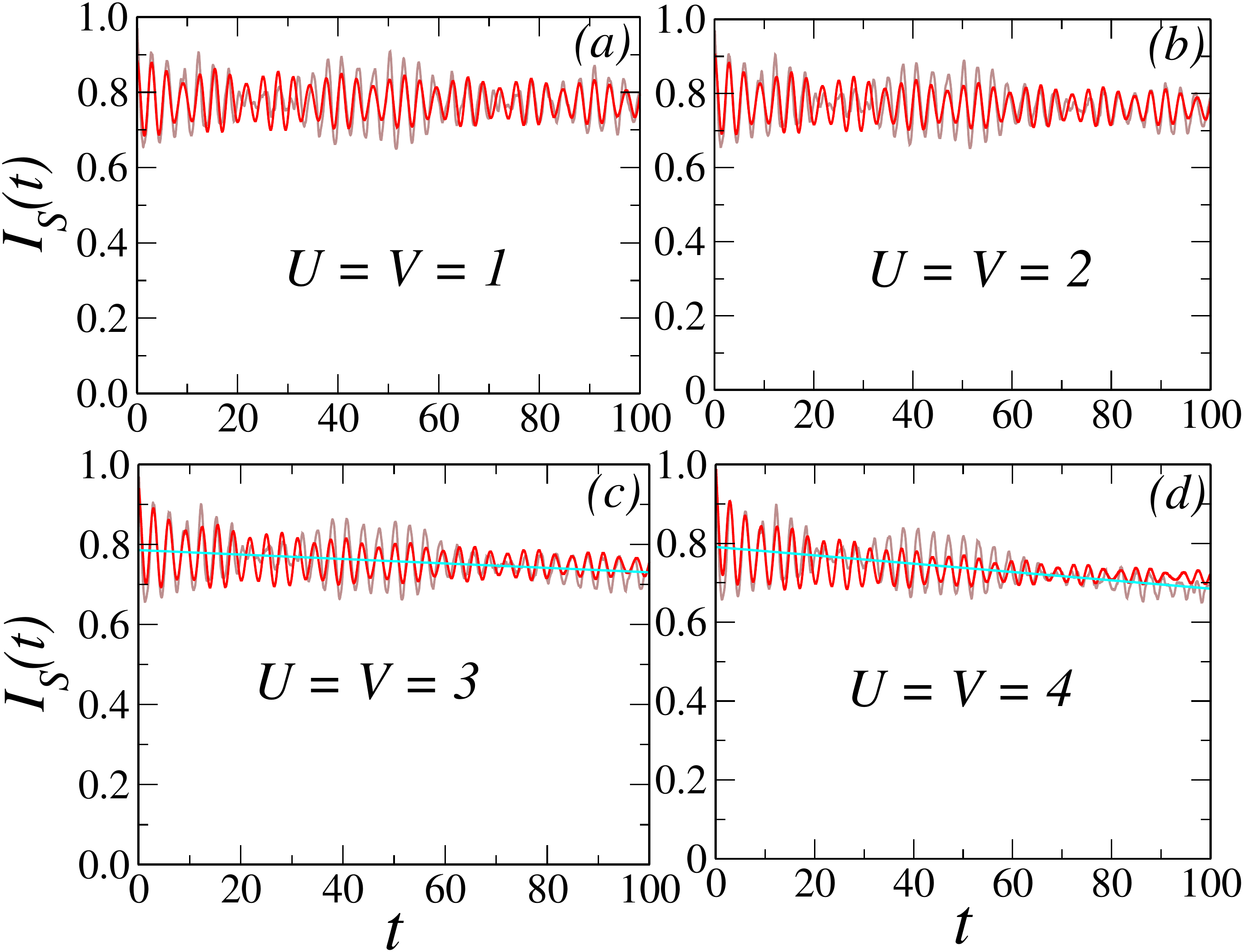}}
\caption{ Spin imbalance $I_S(t)$ vs time $t$ (in units of $\hbar/t_{hop}$) working at $\Delta=12$, 
and for different interaction strength when $U=V$ (where $V$ is the prefactor of long-range interaction and $U$ is on-site interaction):
(a) $U=V=1$, (b) $U=V=2$, (c) $U=V=3$, and (c) $U=V=4$.
The red curves represent the power-law fit to $I_S(t)$ using the fitting function described in the text, while the cyan curve represents a linear fit to $I_S(t)$.} 
\label{fig12}
\end{figure}

In  Fig.~\ref{fig12}, we present the averaged spin imbalance $I_S(t)$  at fixed $\Delta=12$ and for different values of interactions strength
 when $U=V$. We find that for lower values of $U=V$, the spin imbalance remains almost constant with time [for $U=V=1$ (Fig.~\ref{fig12}(a)) and 
$U=V=2$  (Fig.~\ref{fig12}(b))]. However, increasing the interaction strength the spin imbalance decays 
{\it linearly} with a very small slope [for  $U=V=3$ (Fig.~\ref{fig12}(c))  and $U=V=4$ (Fig.~\ref{fig12}(d))]   
In the strong disorder limit, the charge degree of freedom freezes and only the spin degree of freedom plays an important role in the dynamics. Thus, 
based on the studies of spin dynamics in Heisenberg models~\cite{punk1,punk2,luitz},
to describe the time evolution of spin imbalance  $I_S(t)$,  we have used the fitting function
$I_S(t)=a e^{-t/\tau} cos\left(\omega_1 t +\theta\right) + b r^{-\eta }+c r^{-\zeta} sin(\omega_2 t+ \theta)$.
The first term of the fitting function captures the fast exponentially decaying oscillations $\omega_1$ and relaxation time $\tau$
similar to the clean case of Heisenberg model~\cite{punk2}. The second term describes the power-law decay (with non-universal exponent $\eta$) of the spin imbalance. 
The third term contains the subdominating power-law decay (exponent $\zeta$) with characteristic oscillations $\omega_2$
of the spin imbalance $I_S(t)$~\cite{luitz}. 

\begin{figure}[h]
\centering
\rotatebox{0}{\includegraphics*[width=\linewidth]{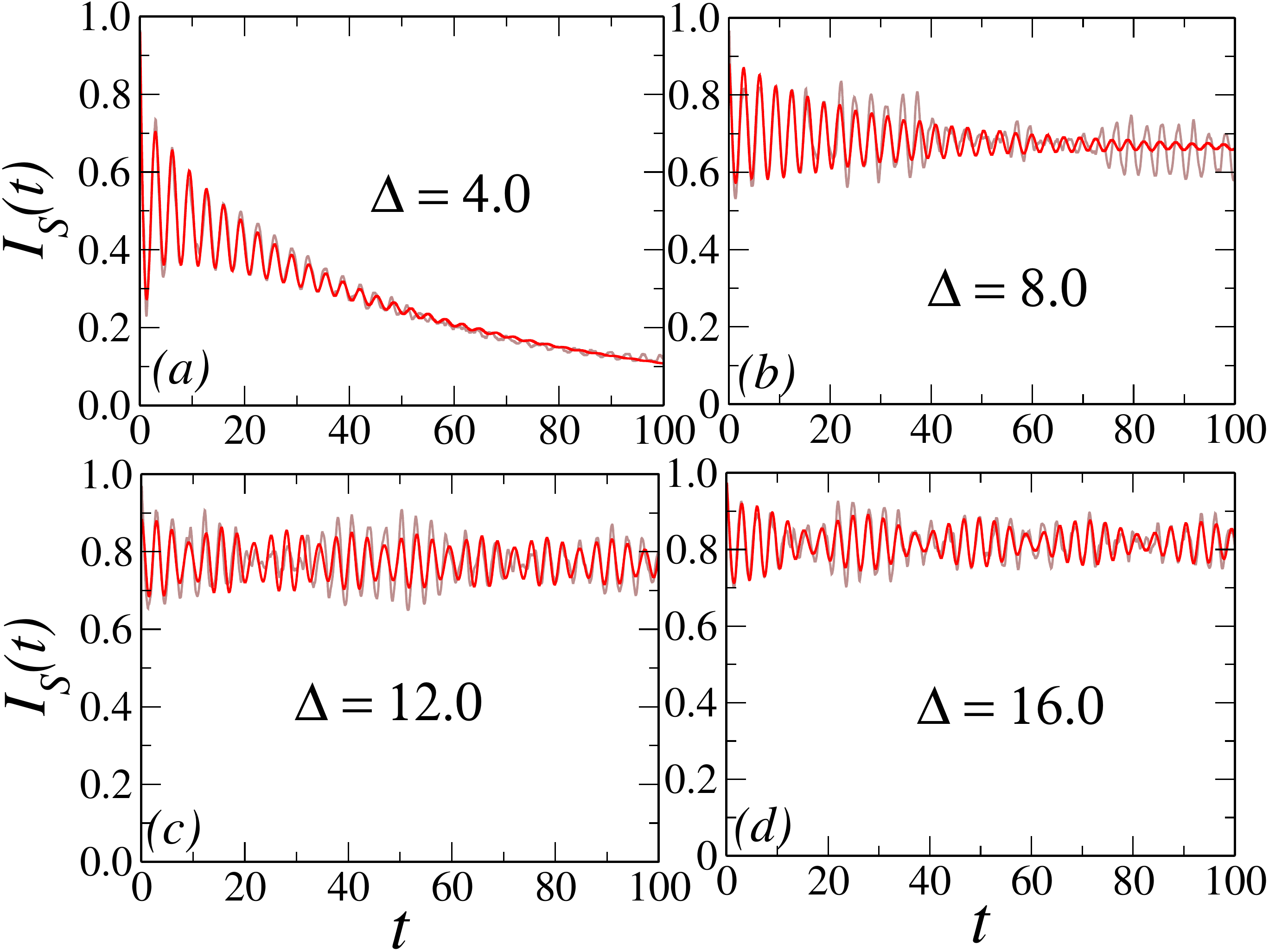}}
\rotatebox{0}{\includegraphics*[width=\linewidth]{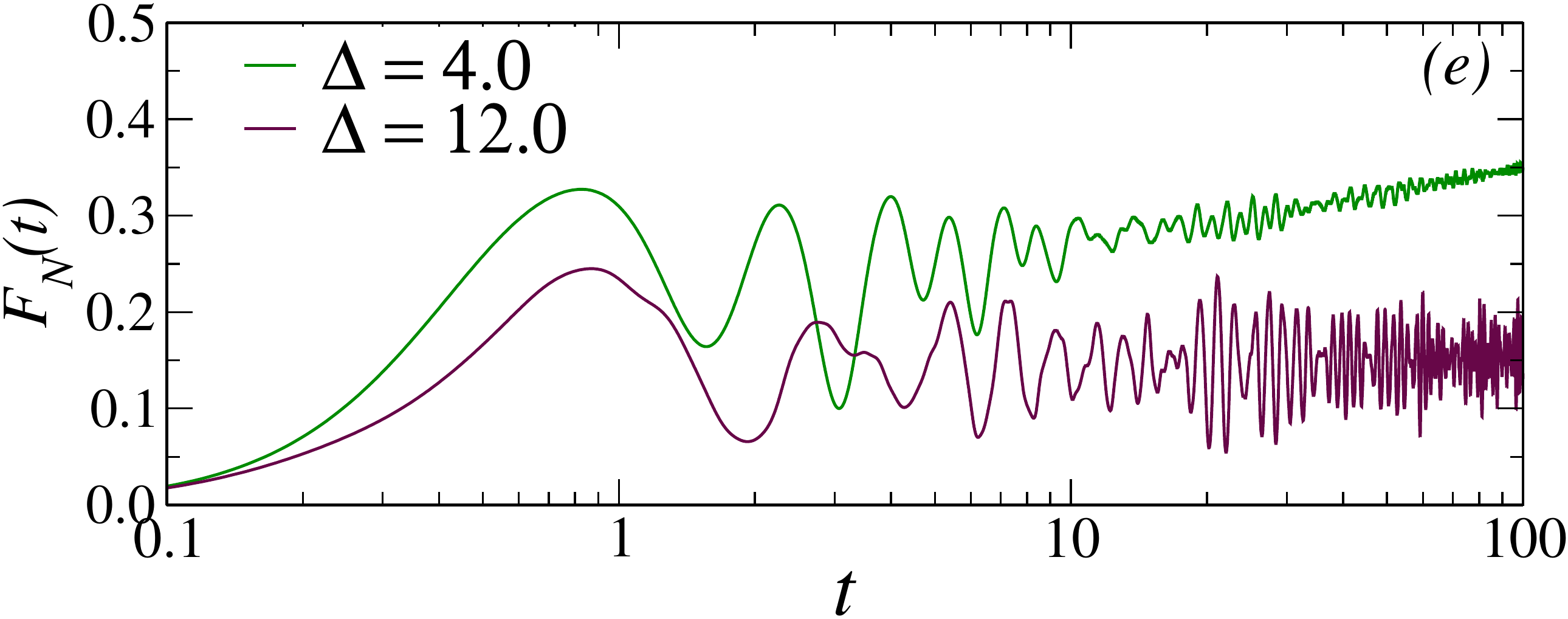}}
\caption{Spin imbalance $I_S(t)$ vs time $t$ (in units of $\hbar/t_{hop}$) working at $U=V=1$, 
and for different values of disorder strength $\Delta$:
(a) $\Delta=4$, (b) $\Delta=8$, (c) $\Delta=12$, and (d) $\Delta=16$. 
The red curve represents the fit to $I_S(t)$ using the fitting function described in the text.
(e) Bipartite charge fluctuation $F_{N}(t)$ vs time $t$} 
\label{fig13}
\end{figure}

Interestingly we find $\omega_1$ is almost independent of interaction strength and 
takes the same value ($\omega_1=1.99$) for all the range $U=V$ considered, while the relaxation time $\tau$ decreases with increasing $U=V$.
The power-law exponent [$\eta=$0.001 ($U=V=1$),  0.008 ($U=V=2$), 0.016 ($U=V=2$), and 0.038 ($U=V=4$)],
 increases with increase in interaction strength, showing a slight increase in the delocalization of spin with increase in interaction strength.
The characteristic oscillation $\omega_2$ of spin-imbalance takes different values depending on interaction strength
[$\omega_2=$2.56 ($U=V=1$), 3.13 ($U=V=2$), 2.5 ($U=V=2$), and 0.9 ($U=V=4$)].  
 As shown in Fig.~\ref{fig12}(d), for larger values of interaction strength $U=V=4$
the decay of spin imbalance with time is better described by the linear fitting function $I_S(t)=0.8-0.001t$.     
Thus, to describe the decay of $I_S(t)$  with time for large values of $U=V=8$ in Fig.~\ref{fig4}(a), we have used the linear fitting function. 

Figure~\ref{fig13} contains the average spin imbalance $I_S(t)$ at $U=V=1$ and for different values of disorder strength $\Delta$.
For low values of disorder $\Delta=4$, the spin imbalance decays with a faster rate. 
The decay of spin imbalance can by described by using an exponential fitting function 
$I_S(t)=a e^{-t/\tau} cos\left(\omega_1 t +\theta\right) + b e^{-\eta t}+c e^{-\zeta t} sin(\omega_2 t+ \theta)$,
with  decay exponent $\eta=0.016$. Whereas, with increase in disorder ($\Delta \gtrsim 8.0$) the time evolution of $I_S(t)$ can be described by 
the power-law fitting function  $I_S(t)=a e^{-t/\tau} cos\left(\omega_1 t +\theta\right) + b r^{-\eta }+c r^{-\zeta} sin(\omega_2 t+ \theta)$.
We find the relaxation time $\tau$ increases with increasing $\Delta$. The power-law exponent $\eta$ decreases with increasing
$\Delta$ [$\eta=$ 0.024 ($\Delta=8$), 0.0025 ($\Delta=12$), and 0.0002 ($\Delta=16$)], showing an increase in localization of the spin with an increase in the
disorder strength. The characteristic oscillation $\omega_2$ of $I_S(t)$ depends on disorder $\Delta$ [$\omega_2=$ 0.77 ($\Delta=$8), 
2.5 ($\Delta=$12), and 2.01 ($\Delta$=16)]. Figure~\ref{fig13}(e) shows bipartite charge fluctuations $F_N(t)$ vs time $t$
for the product state with singlons (N\'eel state).
For lower values of disorder strength $\Delta=4.0$, $F_N(t)$ grows algebraically. On the other hand, for larger values of $\Delta=12$,
 $F_N(t)$ oscillate around a fixed value, indicating localization of the charge degree of freedom~\cite{bar}.

\subsection{Long time dynamics of charge and spin imbalance: With doublons and N\'eel states }
\begin{figure}[h]
\centering
\rotatebox{0}{\includegraphics*[width=\linewidth]{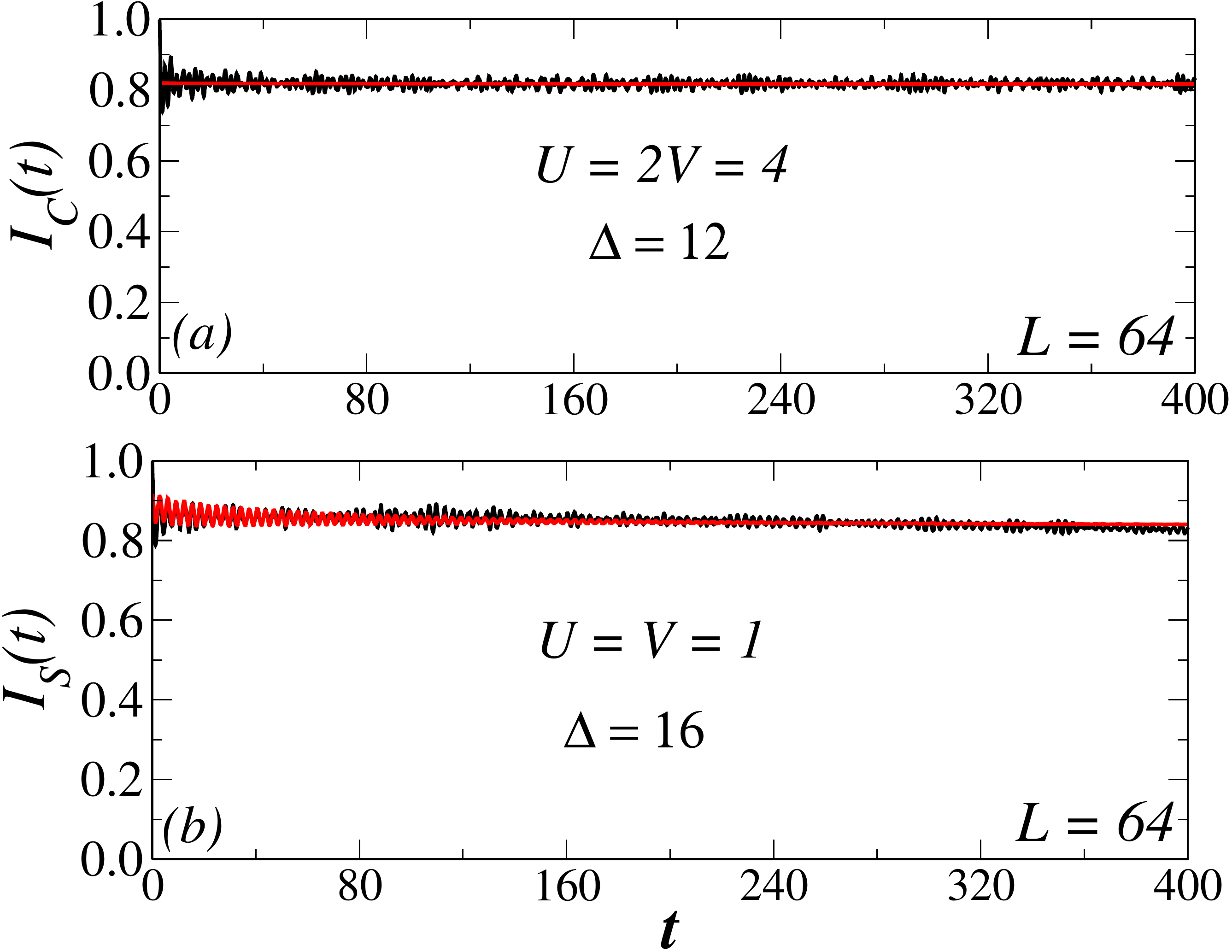}}
\caption{Long time dynamics of charge and spin imbalance using TDVP1 method 
for larger system size $L=64$.(a)The charge imbalance $I_C(t)$ vs time $t$ (in units of $\hbar/t_{hop}$) 
for product state with doublons at $U=2V=4$ and  $\Delta=12$ .
(b) The spin imbalance $I_S(t)$ vs time $t$ for product state with N\'eel state, working at $\Delta=16$
and  $U=1,V=1$.
} 
\label{fig14}
\end{figure}
To study the long-time dynamics, we have carried out the time evolution for significantly
larger system size $L=64$ for product states with doublons and N\'eel states.
Using the time evolution method TDVP1 (with GSE to enlarge the bond-dimension), based on the ITensor library~\cite{itensor},
we are able to reach $t=400 \hbar/t_{hop}$ for large disorder $\Delta$. 
Figure~\ref{fig14}(a) contains the time evolution of average charge imbalance $I_C(t)$, for the
 product state with doublons $|\uparrow \downarrow 0 \uparrow \downarrow 0 \uparrow \downarrow 0 \uparrow \downarrow 0 \uparrow \downarrow \rangle$
 at $U=2V=4$ and  $\Delta=12$. Interestingly we find that the charge imbalance saturates to a constant value and
does not decay with time. The decay exponent $\eta$ obtained by a power-law fit to the charge imbalance $I(t)\sim t^{-\eta}$,
approaches to zero [$\eta=0.0005$], confirming the localization of charge for the product state with doublons. 

Figure~\ref{fig14}(b) plots the time evolution of the average spin-imbalance $I_S(t)$ for the product state with singlons $|\uparrow \ \downarrow  
\ \uparrow \  \downarrow \  \uparrow \  \downarrow \  \uparrow \  \downarrow  \ \uparrow \  \downarrow \ \rangle$ at 
$U=1,V=1$, and $\Delta=16$. We find a very slow decay of the spin imbalance $I_S(t)$ even after long time.
 The decay of $I_S(t)$ with time can be described by the power-law fitting function  
$I_S(t)=a e^{-t/\tau} cos\left(\omega_1 t +\theta\right) + b r^{-\eta }+c r^{-\zeta} sin(\omega_2 t+ \theta)$,
where $\eta=0.009$ shows the very slow decay of $I_S(t)$ with time even for significantly
larger system  $L=64$ for cold atom experiments. However, there is a possibility 
that due to the spin $SU(2)$ symmetry of the system,
 the spin imbalance may decay after very long time. 


\begin{thebibliography}{99}
\bibitem{caveat-appendix} However, for a number of sites $L>2$ of fermions, the calculations of the effective exchange constant $J_{r_i,r_j}$ will have higher order contributions (i.e, exchange from distant sites) and from long-range interaction $V/r^3$. 
Thus, in order to obtain the  effective exchange constant for $L$ sites fermions, addition theoretical work is required.
\end{thebibliography}
\end{document}